\documentclass[%
a4paper,
superscriptaddress,
notitlepage,
 amsmath,amssymb,
 aps,
 twocolumn,
 prl
 floatfix
]{revtex4-1}

\usepackage{tikz}
\usetikzlibrary{calc,arrows,decorations.pathreplacing,backgrounds}
\colorlet{mylinkcolor}{blue!66!black!80}
\usepackage{mathtools,amssymb,amsfonts}
\usepackage[colorlinks=true,linkcolor=mylinkcolor,citecolor=mylinkcolor,filecolor=cyan,urlcolor=mylinkcolor,breaklinks=true]{hyperref}
\usepackage{xcolor}

\usepackage[utf8]{inputenc}
\usepackage{booktabs}
\usepackage{bbm,mathrsfs,mathtools}
\usepackage{graphicx}
\usepackage{hyperref}

\newcommand{\e}[1]{\mathrm{e}^{#1}}

\DeclareMathOperator{\sech}{sech}

\DeclareSymbolFontAlphabet{\mathscrsfs}{rsfs}

\newtheorem{theorem}{Theorem}

\begin{document} 
\title{Local Order Controls the Onset of Oscillations in the Nonreciprocal Ising Model}

\author{Kristian Blom}
\affiliation{Mathematical bioPhysics group, Max Planck Institute for Multidisciplinary Sciences, G\"{o}ttingen 37077, Germany}

\author{Uwe Thiele}
\affiliation{Institute of Theoretical Physics, University of M\"{u}nster, M\"{u}nster 48149, Germany}
\affiliation{Center for Nonlinear Science (CeNoS),
University of M\"{u}nster, M\"{u}nster 48149, Germany}
\affiliation{Center for Multiscale Theory and Computation (CMTC),
University of M\"{u}nster, M\"{u}nster 48149, Germany}

\author{Alja\v{z} Godec}
\email{agodec@mpinat.mpg.de}
\affiliation{Mathematical bioPhysics group, Max Planck Institute for Multidisciplinary Sciences, G\"{o}ttingen 37077, Germany}

\date{\today}

\begin{abstract}
We elucidate the
generic bifurcation behavior of local and global order in the nonreciprocal Ising model evolving under Glauber dynamics. We show
that a critical magnitude of
nearest-neighbor correlations within the respective lattices controls
the emergence of coherent oscillations of global
order as a result of frustration. 
Local order is maintained
during these oscillations, implying nontrivial spatiotemporal
correlations. Long-lived states
emerge in the
strong-interaction regime. The residence time in either of these
states eventually diverges, giving rise to ordered non-equilibrium
trapped states and a loss of ergodic behavior via a saddle-node-infinite-period
bifurcation. Our work provides a comprehensive
microscopic understanding of the nonreciprocal Ising model beyond the mean-field approximation. 
\end{abstract}
\maketitle
%---------------------------------------------------
%---------------------------------------------------
\section{Introduction}
%---------------------------------------------------
%---------------------------------------------------
The last decade saw a surge of interest in many-body lattice systems with nonreciprocal interactions
\cite{Guislain_2024A,Guislain_2024B,  Seara_2023, PhysRevE.94.042139, Collet2014, Collet2019,avni2023nonreciprocal,PhysRevLett.130.198301,PhysRevE.109.034131, osat2023non, PhysRevLett.133.028301}. At the microscopic level, nonreciprocal interactions violate Newton's third law and result in broken
detailed balance \cite{PhysRevX.5.011035,
   PhysRevX.14.021014}, thereby driving the system out of equilibrium.  On the collective level, nonreciprocally interacting
systems can resist coarsening and self-organize into
dynamic
states with unique spatiotemporal patterns, including traveling
and oscillatory states \cite{te2022derivation, PhysRevE.103.042602,
  PhysRevX.10.041009, PhysRevLett.95.198101, 10.1093/imamat/hxab026,
  PhysRevE.107.064210, PhysRevE.109.L062602,
  frohoff2023non, GLFT2025prl}. Phenomenologically, such systems are typically described using nonvariational couplings of Allen-Cahn models (for nonconserved dynamics) or Cahn-Hilliard models (for conserved dynamics), corresponding to models A and B, respectively, as outlined in \cite{HoHa1977rmp}.
   
 By introducing two nonreciprocally coupled Ising lattices, various studies have shown under which conditions nonreciprocity induces temporal oscillations in the magnetization \cite{avni2023nonreciprocal, Guislain_2024A, Seara_2023,
  Guislain_2024B, PhysRevE.94.042139, Collet2014, Collet2019}. These works revealed intriguing
phenomena, such as Hopf instabilities
\cite{avni2023nonreciprocal, 
  Guislain_2024A, Guislain_2024B, Collet2019}, saddle-node bifurcations
\cite{LHK2023jpsj, avni2023nonreciprocal}, and hidden collective oscillations
\cite{Guislain_2024B}. However, so far these studies have been limited to phenomenological and mean-field theory, raising the question to what extent these results apply beyond their respective approximations. 

Here, we go beyond mean-field reasoning
and explicitly incorporate nearest-neighbor
correlations into a thermodynamically consistent description of two nonreciprocally coupled Ising models
We consider both \emph{global and local order}, and explain why a critical magnitude of
nearest-neighbor correlations controls the symmetry-breaking
transition in the global order, in turn bounding the onset of
coherent oscillations.  
We elucidate how the bifurcation behavior depends on
the interaction strength and highlight stark differences
in the spatiotemporal dynamics of all-to-all (mean-field) 
versus short-range-interacting systems; 
the square and Bethe lattices display
equivalent behavior that is strikingly different from the
all-to-all lattice. 
%---------------------------------------------------
%---------------------------------------------------
\section{Model}
%---------------------------------------------------
%---------------------------------------------------
Consider a pair of
lattices denoted by $\mu=a,b$, as shown in Fig.~\ref{Fig1}a,
each having a coordination
number $z$ and periodic boundary conditions. On each lattice there are $N$ spins that can assume two states $\sigma^{\mu}_{i}=
\pm1$, with $i\in \{1,...,N \}$ enumerating the spin's location.
Each spin interacts with its $z$ nearest neighbors on the same
lattice and the spin at the equivalent position on the opposing
lattice. The \emph{local} interaction \footnote{The local interaction is not equal to the total energy of the system, since the interaction energy for the two lattices $a$ and $b$ is different.} energy of spin $i$ on lattice $\mu$ can be written as 
\begin{equation}
    E^{\mu}_{i}=-J_{\mu}\sigma^{\mu}_{i}\sum_{\langle i|j \rangle}\sigma^{\mu}_{j}-K_{\mu}\sigma^{a}_{i}\sigma^{b}_{i},
    \label{E}
\end{equation}
where $\langle i|j \rangle$ denotes a sum over nearest-neighbors $j$
of spin $i$. Throughout, we express energies in
units of
$k_{\rm B}T$, where $T$ is the temperature of the heat bath. The parameter
$J_{\mu}$ denotes the coupling within
lattice $\mu$, and $K_{\mu}$ denotes the (directed) coupling between the spins in $\mu$ and those of the opposing lattice,  
respectively. When $K_{a} \neq K_{b}$, equivalent
spins on the opposing lattices interact nonreciprocally.

We consider single spin-flip Glauber dynamics
\cite{glauber_timedependent_1963}.~Let $P(\boldsymbol{\sigma};t)$ be
the probability at time $t$ to find the system in state
$\boldsymbol{\sigma}=\{\sigma_{1}^{a},\sigma_{1}^{b},...,\sigma_{N}^{a},\sigma_{N}^{b}\}$
which is governed by the master equation
\begin{equation}
\frac{{\rm d}P(\boldsymbol{\sigma};t)}{{\rm d}t}=\sum_{\mu, i}w^{\mu}_{i}(-\sigma^{\mu}_{i})P(\boldsymbol{\sigma}^{\prime}_{\mu,i};t){-}
   w^{\mu}_{i}(\sigma^{\mu}_{i})P(\boldsymbol{\sigma};t),
   \label{masterequation}
\end{equation}
where
$\boldsymbol{\sigma}^{\prime}_{\mu,i}=\{\sigma_{1}^{a},\sigma_{1}^{b},...,-\sigma^{\mu}_{i},...,\sigma_{N}^{a},\sigma_{N}^{b}\}$
is a state which differs from state $\boldsymbol{\sigma}$ by one
spin flip.  
The transition rates $w^{\mu}_{i}(\sigma^{\mu}_{i})$ to flip a spin $\sigma^{\mu}_{i}\rightarrow -\sigma^{\mu}_{i}$ are uniquely specified by limiting
the interactions to nearest neighbors, imposing isotropy in
position space, and requiring that for $K_{a}=K_{b}$ the transition
rates obey detailed balance.  These physical restrictions then lead to the general result \cite{Guislain_2024A, Guislain_2024B}
\begin{equation}
    w^{\mu}_{i}(\sigma^{\mu}_{i})=\left[1-\tanh{\left(\Delta E^{\mu}_{i}/2\right)}\right]/2\tau,
\end{equation}
where $\Delta E^{\mu}_{i}=-2E^{\mu}_{i}$ is the change in energy on
the $\mu = a,b$ lattice after spin conversion
$\sigma^{\mu}_{i}\rightarrow -\sigma^{\mu}_{i}$, and $\tau$
is an intrinsic timescale to attempt a single spin-flip. 
%---------------------------------------------------
%---------------------------------------------------
\subsection{Global and local order parameters}
%---------------------------------------------------
%---------------------------------------------------
We are interested in the temporal dynamics of global and local order parameters averaged over all spins. The magnetization or global order \cite{Saito1976} is given by 
\begin{equation}
    m^{\mu}(t) \equiv\frac{1}{N}\sum_{i=1}^{N}\langle \sigma^{\mu}_{i}(t) \rangle \ \in [-1,1],
    \label{mmu}
\end{equation}
where $\langle f(t) \rangle \equiv
\sum_{\boldsymbol{\sigma}}P(\boldsymbol{\sigma};t)f(\boldsymbol{\sigma})$. The three local order parameters \cite{Saito1976} are
\begin{align}
     q^{\mu\mu}(t) &\equiv \frac{2}{zN} \sum_{i=1}^{N}\sum_{\langle i|j \rangle }\langle \sigma_{i}^{\mu}(t)\sigma_{j}^{\mu}(t) \rangle \ \in [-1,1], \label{qmumu}\\
    q^{ab}(t) &\equiv \frac{1}{N}
    \sum_{i=1}^{N}\langle
      \sigma_{i}^{a}(t)\sigma_{i}^{b}(t) \rangle \ \ \ \ \ \ \ \ \in [-1,1],
    \label{qmunu}
\end{align}
with correlations 
\begin{equation}
\mathcal{C}^{\mu\nu}(t)\equiv q^{\mu\nu}(t)-m^{\mu}(t)m^{\nu}(t).
\label{Cmunu}
\end{equation}
We distinguish between the local order within and
between the
lattices. The alignment of spin pairs within lattice
$\mu$ is quantified by $q^{\mu\mu}(t)$, and $q^{ab}(t)$ (also known as the overlap \cite{PhysRevB.109.184203, PhysRevLett.130.207102})
measures the alignment of equivalent spins between both
lattices. The normalization in Eq.~\eqref{qmumu} arises because $zN/2$ is the number of nearest neighbor pairs in a
periodic lattice with coordination number $z$.
%---------------------------------------------------
%---------------------------------------------------
\subsection{Evolution equations beyond the mean-field approximation}
%---------------------------------------------------
%---------------------------------------------------
Our first main result is an
\emph{exact} set of coupled differential equations for the order parameters in the thermodynamic limit $N\rightarrow\infty$, which reads (see Appendix \ref{AppendixA} for a detailed derivation)
\begin{align}
    \!\!\!\!\tau\frac{{\rm d}m^{\mu}(t) }{{\rm d}t}{+}m^{\mu}(t)
   &{=}\sum_{l,n}\mathcal{P}^{\mu}_{l,n}(t)\tanh(U^{\mu}_{l,n}), \label{Eq1} \\
   \!\!\!\!\tau\frac{{\rm d}q^{\mu\mu}(t)}{{\rm d}t}{+}2q^{\mu\mu}(t)
   &{=}\frac{2}{z}\!\sum_{l,n}(2l{-}z)\mathcal{P}^{\mu}_{l,n}(t)\tanh(U^{\mu}_{l,n}), \label{Eq2} \\ 
   \!\!\!\!\tau\frac{{\rm d}q^{ab}(t)}{{\rm d}t}{+}2q^{ab}(t) 
   &{=}\!\sum_{\mu}\!\sum_{l,n}(2n{-}1)\mathcal{P}^{\mu}_{l,n}(t)\!\tanh(U^{\mu}_{l,n}), \label{Eq3}
\end{align}
where $\sum_{l,n}\equiv \sum_{l=0}^{z}\sum_{n=0}^{1}$ is a sum over all possible values of neighboring up spins on the same ($l\in\{0,\ldots,z\}$) and opposing ($n\in\{0,1\}$) lattice, and 
\begin{equation}
    U^{\mu}_{l,n}\equiv [2l-z]J_{\mu}+[2n-1]K_{\mu}
    \label{U}
\end{equation}
parameterizes the change in energy upon flipping a spin with such a local environment. Finally,
$\mathcal{P}^{\mu}_{l,n}(t)\in[0,1]$ is the time-dependent probability of selecting an
up or down spin that has $l$ neighboring up spins on the same
lattice and $n$ neighboring up spins on the opposing lattice.
The probability is normalized as
\begin{equation}
    \sum_{l,n}\mathcal{P}^{\mu}_{l,n}(t)=1.
\end{equation}
Equations~\eqref{Eq1}-\eqref{Eq3} are
not yet closed; evaluating $\mathcal{P}^{\mu}_{l,n}(t)$ for an arbitrary
lattice is a daunting combinatorial task, as it depends on
microscopic details and, therefore, on an infinite hierarchy of
order parameters. However, we can approximate
$\mathcal{P}^{\mu}_{l,n}(t)$ to different levels of accuracy, which we do in the next section. 

Note that evolution equations for the nonreciprocal Ising model on the fully connected mean-field lattice
have been constructed in \cite{avni2023nonreciprocal}, and are
reported in Appendix \ref{AppendixMF} for completeness. 
\begin{figure*}
    \centering
    \includegraphics[width=\textwidth]{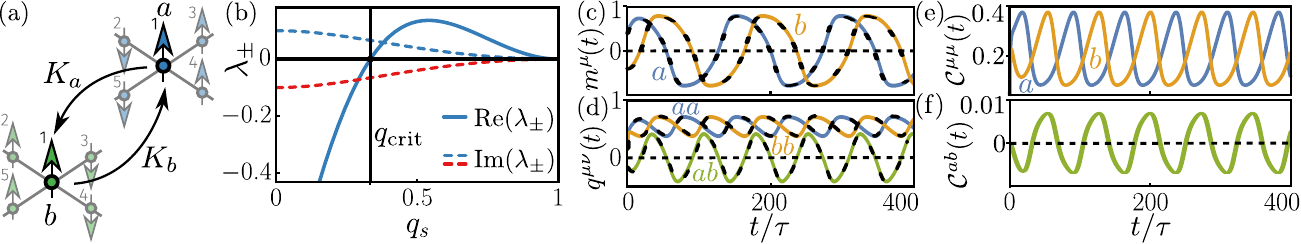}
    \caption{(a)~Schematic of two Ising lattices $a$ and $b$ with coordination number $z{=}4$ and
    cross-coupling $(K_{a},K_{b})$. For $K_{a}\neq
    K_{b}$ the cross-coupling is nonreciprocal.~(b)~Eigenvalues
    $\lambda_{\pm}$ of the linear stability matrix [Eq.~\eqref{BGeigval}] as a function of the steady-state
    local order $q_{s}$ for $K_{a}=-K_{b}=0.1$.~When
    $q_{s}\geq q_{\rm crit}$ [black vertical line;~Eq.~\eqref{qcrit}] the
    real parts of $\lambda_{\pm}$
    are non-negative, resulting in coherent 
    oscillations.~(c-f)~Temporal evolution of the magnetization
    [Eq.~\eqref{mmu}] (c), local order [Eqs.~\eqref{qmumu}-\eqref{qmunu}] (d), and local correlations [Eq.~\eqref{Cmunu}] (e,f) for $K_{a}=-K_{b}=K=0.1$ and $J_{a}=J_{b}=0.4$. Black dashed lines in (c,d) are obtained with Monte-Carlo simulations on the Bethe lattice (see Appendix \ref{AppendixMC} for details).~In all panels we consider $z=4$.}
    \label{Fig1}
\end{figure*}
%---------------------------------------------------
%---------------------------------------------------
\subsection{Bethe-Guggenheim approximation}
%---------------------------------------------------
%---------------------------------------------------
An accurate closed-form expression for $\mathcal{P}^{\mu}_{l,n}(t)$ can
be obtained with the Bethe-Guggenheim (BG) approximation (or pair approximation), where we
assume perfect mixing of nearest neighbor spin pairs. This
approximation is exact on loopless lattices such as the Bethe lattice
\cite{blom2023pair}, or large random graphs with fixed coordination
number \cite{D_A_Johnston_1998SM, Deepak_Dhar_1997SM}. As we show
in Sec.~\ref{SecV}, the spatiotemporal dynamics on these lattices
agrees, in contrast to that on a mean-field lattice, with the behavior on the
square lattice.  We split $\mathcal{P}^{\mu}_{l,n}(t)$ in ``up'' and ``down'' spin contributions 
\begin{equation}
    \mathcal{P}^{\mu}_{l,n}(t)=\mathcal{P}^{\mu+}_{l,n}(t)+\mathcal{P}^{\mu-}_{l,n}(t),
\end{equation}
where, e.g., $\mathcal{P}^{\mu+}_{l,n}(t)\in[0,1]$ is the probability of
flipping an up spin with $l$ up neighbors on the same and $n$ up neighbors
on the opposing lattice, respectively. These probabilities are derived in Appendix \ref{AppendixB} and read (omitting the explicit $t$-dependence on the right-hand side)
\begin{equation}
    \mathcal{P}^{a\pm}_{l,n}(t)=\frac{C^{z}_{l}(1{\pm}2m^{a}{+}q^{aa})^{\delta^{\pm}_{l}}(1{\pm}m^{a}{-}m^{b}{\mp}q^{ab})^{1-n}}{(1{\pm}m^{a})^{z}(1{-}q^{aa})^{-\delta^{\mp}_{l}}(1{\pm}m^{a}{+}m^{b}{\pm}q^{ab})^{-n}}, 
    \label{BG3}
\end{equation}
where $\delta^{+}_{l}=l$, $\delta^{-}_{l}=z-l$, and
\begin{equation}
    C^{z}_{l}\equiv\frac{1}{2^{z+2}}\binom{z}{l}.
\end{equation}
 The expression for
$\mathcal{P}^{b \pm}_{l,n}(t)$ follows from Eq.~\eqref{BG3} by
interchanging $m^{a} \leftrightarrow m^{b}$ and
$q^{aa}\leftrightarrow q^{bb}$. Inserting
Eq.~\eqref{BG3} into Eqs.~\eqref{Eq1}-\eqref{Eq3} yields a closed system of five coupled
nonlinear differential equations.   
%---------------------------------------------------
%---------------------------------------------------
\section{Linear analysis and the Hopf bifurcation}
%---------------------------------------------------
%---------------------------------------------------
We focus
on the symmetric nonreciprocal setting $J_{a}=J_{b}=J$ and
$K_{a}=-K_{b}=K$, also known as the perfectly nonreciprocal
\cite{PhysRevX.14.011029} setting, while keeping $z$ general. We start with the
linear stability analysis of Eqs.~\eqref{Eq1}-\eqref{Eq3} around their
steady state. The trivial steady state is given by $m^{\mu}_{s}=0$, $q^{ab}_{s}=0$, and
\begin{equation}
    q^{\mu\mu}_{s}\equiv q_{s}(J,K),
\end{equation}
which is explicitly given in Appendix \ref{AppendixQ} for various values of $z$. Up to first order, small perturbations $\delta \mathbf{m}(t)\equiv(\delta m^{a}(t),\delta m^{b}(t))$ decouple from perturbations $\delta \mathbf{q}(t)\equiv(\delta q^{aa}(t),\delta q^{bb}(t),\delta q^{ab}(t))$, and we obtain the linear equation 
\begin{equation}
\tau \frac{{\rm d}\delta \mathbf{m}(t)}{{\rm d}t}{=} \begin{pmatrix}
M_{1}(q_s;J,K) & -M_{2}(q_s;J,K) \\
M_{2}(q_s;J,K) & M_{1}(q_s;J,K)
\end{pmatrix}
\delta \mathbf{m}(t).
\end{equation}
The linear equation for $\delta \mathbf{q}(t)$ is given in \footnote{See Supplementary Material at [...]} and does not play any further role here up to linear order. 
The elements of the linear stability matrix read
\begin{align}
    \!\!M_{1}(q_s;J,K)&=\frac{q_{s}/q_{\rm crit}-1}{1+q_{s}}[1-2\sum\nolimits_{l,n} \!\!\overline{\mathcal{P}}^{+}_{l}\tanh(U^{a}_{l,n})], \nonumber \\
   \!\!M_{2}(q_s;J,K)&=\sum\nolimits_{l,n}\!(2n{-}1)[\overline{\mathcal{P}}^{+}_{l}{+}\overline{\mathcal{P}}^{-}_{l}] \tanh(U^{a}_{l,n}), \label{d1}
\end{align}
where 
\begin{equation}
    \overline{\mathcal{P}}^{\pm}_{l}(q_{s})\equiv \mathcal{C}^{z}_{l}(1\mp q_{s})^{z-l}(1 \pm q_{s})^{l}
    \label{overlineP}
\end{equation} 
are the probabilities \eqref{BG3} evaluated at steady-state values, and we introduced the critical local order
\begin{equation}
    q_{\rm crit}\equiv \frac{1}{z-1},
    \label{qcrit}
\end{equation}
which \emph{only} depends on the coordination number of the lattice, and sets a critical value for the steady-state local order. The solution of the linear stability equation can be written as
\begin{equation}
    \mathbf{\delta m}(t)=\sum_{k=\pm}\mathcal{A}_{k}\e{\lambda_{k}t/\tau}\boldsymbol{\nu}_{k},
\end{equation}
where
$\mathcal{A}_{\pm}$ are set by the initial conditions,
$\boldsymbol{\nu}_{\pm}=(\mp {\rm i},1)^{\rm T}$ are the eigenvectors of the linear stability matrix, and $\lambda_{\pm}(q_s;J,K)$ the corresponding eigenvalues
\begin{equation}
    \lambda_{\pm}(q_s;J,K)=M_{1}(q_s;J,K) \pm {\rm i}M_{2}(q_s;J,K),
    \label{BGeigval}
\end{equation}
$\rm i$ being the imaginary unit.~Since ${M_{2}(q_s;J,K\neq0)\neq0}$ (see
proof in Appendix \ref{AppendixD}), the eigenvalues are complex for $K\neq0$ (see dashed
lines in Fig.~\ref{Fig1}b), resulting in oscillatory
perturbations.~The Hopf bifurcation \cite{strogatz2018nonlinear},
also called type-${\rm II}_{\rm o}$ instability
\cite{RevModPhys.65.851}, occurs when
complex conjugate
eigenvalues transit
the imaginary axis in the complex
plane.~According to Eq.~\eqref{BGeigval} this occurs when
$M_{1}(q_s;J,K)=0$ implying
${q_{s}(J,K)=q_{\rm crit}}$ as seen from Eq.~\eqref{d1}.~The Hopf bifurcation is thus
set by the critical value $q_{\rm crit}$ for
local
order, and for $q_{s}> q_{\rm crit}$ we have ${\rm
  Re}(\lambda_{\pm})>0$ (solid line in Fig.~\ref{Fig1}b).~In other
words, when
spins on the
respective lattices are sufficiently aligned, a transition
to an oscillatory state occurs
as shown in Fig.~\ref{Fig1}c,d.

The critical value of the local order that determines the onset of coherent oscillations is our
second main result
that generalizes
to other approximation
schemes beyond the mean-field approximation (see Appendix \ref{AppendixMon}). After sufficient local order is attained within the
lattices, the frustration due to the nonreciprocal coupling gives
rise to
coherent oscillations: for $K>0$ a spin $\sigma^{a}_{i}$ wants to align
with $\sigma^{b}_{i}$ that, in turn, tends to
misalign with $\sigma^{a}_{i}$. This dynamical frustration results in an oscillatory
motion of the order parameters \cite{PhysRevX.14.011029}. 
Notably, for the one-dimensional lattice ($z=2$) we find $q_{\rm
  crit}=1$, which, in contrast to the mean-field prediction
\cite{avni2023nonreciprocal} (see also Appendix \ref{AppendixMF}), correctly implies the nonexistence of a Hopf
bifurcation. 
\begin{figure*}[t!]
    \centering
    \includegraphics[width=\textwidth]{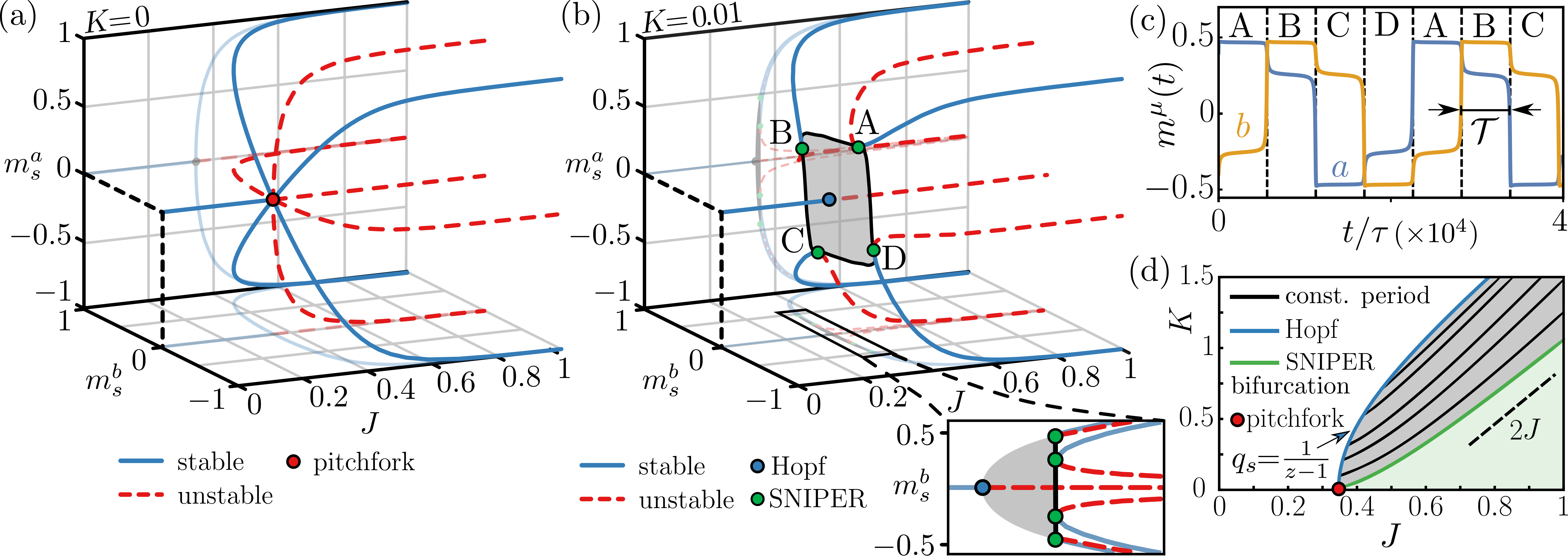}
    \caption{(a,b)~Bifurcation diagram of the magnetization
      $(m^{a}_{s},m^{b}_{s})$ in the absence (a; $K{=}0$) and
      presence (b; $K{=}0.01$) of nonreciprocal coupling as a function
      of $J$.~Inset of (b): Magnification of the bifurcation diagram
      around the Hopf bifurcation projected onto the $(m^{b}_{s},J)$-plane (see black box).~(c) Temporal evolution of the magnetization close to the degenerate
      saddle-node-infinite-period (SNIPER) bifurcation at $(J,K){=}(0.35991,0.01)$. The magnetization
      oscillates coherently between $4$ ghost states ${\rm A}{\rightarrow} {\rm B} {\rightarrow} {\rm C} {\rightarrow} {\rm D}$
      (for $K{<}0$ the direction is reversed), which eventually
      terminate in the
      $4$ respective stable branches at and beyond the SNIPER
      bifurcation. The life-time $\mathcal{T}$ in these ghost states diverges
      upon approaching the SNIPER bifurcation at $J_{\rm SNP}(K)$ (see Fig.~\ref{Fig3}).~(d) Phase diagram of
      $m^{\mu}_{s}$:~the gray region depicts the regime of coherent
      oscillations, with the
      black lines indicating isolines with fixed oscillation period, and the blue line indicating the Hopf bifurcations where $q_{s}=q_{\rm crit}$ [Eq.~\eqref{qcrit}];
      in the light green region
      the magnetization is stationary and nonzero. In all panels we consider $z=4$.}
    \label{Fig2}
\end{figure*}
%---------------------------------------------------
%---------------------------------------------------
\section{Nonlinear analysis and the SNIPER bifurcation}
%---------------------------------------------------
%---------------------------------------------------
Going beyond linear stability, we perform a nonlinear analysis of
Eqs.~\eqref{Eq1}-\eqref{Eq3} through numerical continuation
\cite{10.1145/779359.779362}. 
The resulting bifurcation diagrams are shown in Fig.~\ref{Fig2}(a,b)
and the complete phase diagram in Fig.~\ref{Fig2}(d), which we now explain in detail. 

We start with the non-interacting case with $K=0$ (see
Fig.~\ref{Fig2}a). For small values of $J$, there exists only one
(trivial) stable steady state with $m^{\mu}_{s}=0$, as explained in the
previous section. Increasing $J$ to $\ln{(z/(z-2))}/2$ we find a
pitchfork bifurcation (red dot in Fig.~\ref{Fig2}a), which coincides
with $q_{s}(J,0)=q_{\rm crit}$, and beyond which the trivial stable state becomes unstable. At the pitchfork bifurcation, $4$
stable branches (blue lines in Fig.~\ref{Fig2}a) and $4$ unstable
branches (red dashed lines in Fig.~\ref{Fig2}a) emerge. Unstable branches have zero magnetization in one of the lattices, while stable branches exhibit nonzero equilibrium
magnetization because of spontaneously broken symmetry in both lattices.
\begin{figure}[b!]
    \centering
    \includegraphics[width=0.75\linewidth]{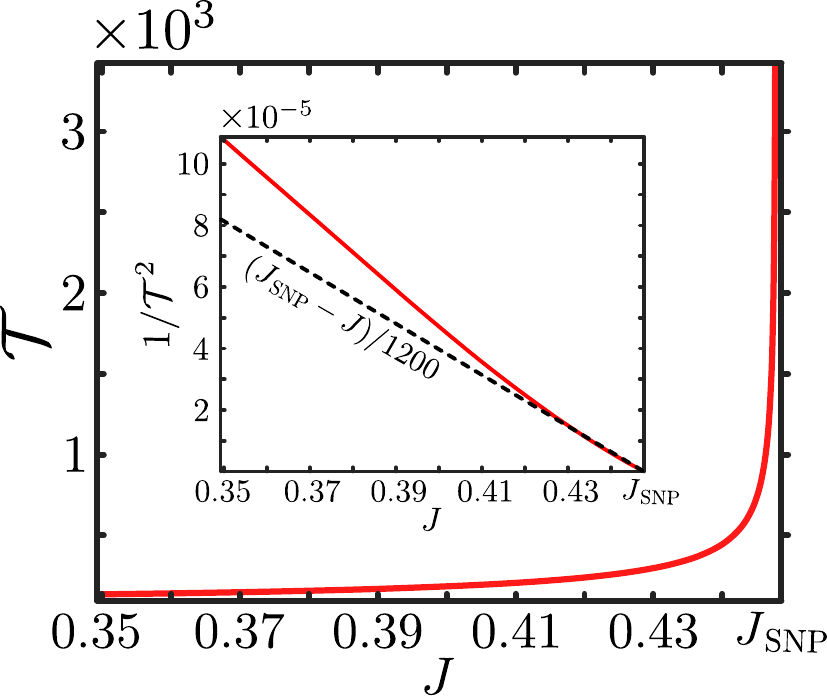}
    \caption{Algebraic divergence of the residence time $\mathcal{T}$ within a ghost state close to the degenerate saddle-node-infinite-period (SNIPER) bifurcation. Approaching the SNIPER bifurcation from below, the residence time diverges algebraically according to Eq.~\eqref{divergence}. Here, we have $K=0.1$ and $z=4$, for which the SNIPER bifurcation occurs at $J_{\rm SNP}\approx0.4477$.  Results are obtained with the continuation package MATCONT \cite{10.1145/779359.779362}.}
    \label{Fig3}
\end{figure}
\begin{figure*}[t!]
    \centering
    \includegraphics[width=\textwidth]{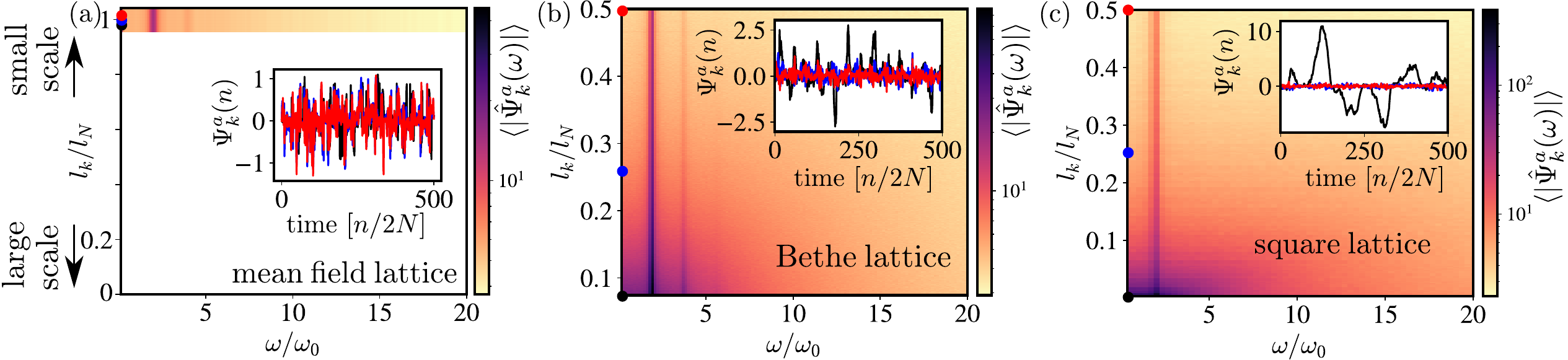}
    \caption{(a-c) Spectral density $\langle
      |\hat{\Psi}^{a}_{k}(\omega)| \rangle$ of modes of coherent oscillations
      [Eq.~\eqref{aomega}] on the (a) mean-field, (b)
      Bethe, and (c) square lattice (see \cite{Note2} for animations of the dynamics on the lattices).~Note that the nonzero eigenvalues of the mean-field lattice are degenerate, which explains the white region in (a).~The values for
      $J$ are chosen in the respective oscillatory regimes (see Appendix \ref{AppendixMC}
      for details), and $K=0.3$ for all three lattices.
      $\langle
      |\hat{\Psi}^{a}_{k}(\omega)| \rangle$ is averaged over
      $500$ independent trajectories, and $\omega_{0}={\rm
        argmax}\langle |\hat{\Psi}^{a}_{1}(\omega)| \rangle $ is the
      natural oscillation frequency.~Resonances are visible at
      multiplies of $2\omega_{0}$, since we ignore the sign of projections.~Insets:~Temporal development of the projected
      microscopic states onto the eigenvectors of the 
      Laplacian matrix of respective lattices.~Colors denote
      three selected eigenvalues (and thus spatial scales) indicated
      by the dots in the main plot.}
    \label{Fig4}
\end{figure*}

Upon setting $K\neq 0$, the pitchfork bifurcation turns into a Hopf
bifurcation (blue dot in Fig.~\ref{Fig2}b), whose $J$-value depends on
$K$ through the relation $q_{s}(J,K)=q_{\rm crit}$ (blue line in
Fig.~\ref{Fig2}d; see Appendix \ref{AppendixQ} for explicit results). Increasing $J$ beyond
the Hopf bifurcation, there is a regime with coherent oscillations
(gray area in Fig.~\ref{Fig2}b,d). In Fig.~\ref{Fig2}d we identify the
isolines of fixed period of the oscillations (black lines). Upon
further increasing $J$ at fixed $K$, we observe a nonlinear
transition from coherent oscillations to a nonzero stationary
magnetization (light green region in Fig.~\ref{Fig2}d),  which is set
by a degenerate saddle-node-infinite-period (SNIPER) bifurcation (green dots in
Fig.~\ref{Fig2}b).
Approaching the SNIPER bifurcation from below, the magnetization in
both lattices starts to oscillate between four long-lived ghost states (see
Fig.~\ref{Fig2}c). These $4$ long-lived states correspond to the virtual configuration, in which
the respective lattices exert a quasi-static magnetic field on each other, and are characterized
by a critical slowing down of the dynamics in the vicinity of the impending SNIPER bifurcation. The residence time
within
these
ghost states
diverges algebraically as (see Fig.~\ref{Fig3}) 
\begin{equation}
\mathcal{T}\propto (J_{\rm SNP}(K)-J)^{-1/2},
\label{divergence}
\end{equation}
where $J_{\rm SNP}(K)$ is the $J$-value of the SNIPER bifurcation at a
given $K$ (green line in Fig.~\ref{Fig2}d). At the SNIPER bifurcation,
one unstable and stable branch emerges, resulting in $4$ unstable
branches (red dashed lines in Fig.~\ref{Fig2}b) and $4$ stable
branches (blue lines in Fig.~\ref{Fig2}b). 
Note that these stable states
with nonzero stationary magnetization also exist on the finite square lattice system (see proof in \cite{Note2}).

The bifurcation diagram obtained with the mean-field approximation has
similar qualitative features as in Fig.~\ref{Fig2}a,b (i.e, a Hopf and SNIPER bifurcation), however, the phase diagram 
 displays a constant Hopf line at a fixed
$J$, independent of $K$ (see \cite{avni2023nonreciprocal} and Fig.~\ref{FigA1}). In Fig.~\ref{Fig2}d we see that the Hopf line
with the BG approximation is $K$ dependent, closely resembling the empirical phase diagram on the qubic lattice (see
\cite{avni2023nonreciprocal}).
%---------------------------------------------------
%---------------------------------------------------
\section{Spatiotemporal dynamics}\label{SecV}
%---------------------------------------------------
%---------------------------------------------------
The results in Fig.~\ref{Fig1}d-f
reveal a high degree of local order in the coherent oscillatory regime. To systematically analyze
spatiotemporal patterns in states with coherent oscillations, we
performed discrete-time Monte Carlo simulations of the nonreciprocal Ising
system with
$2\times N\approx 3\times 10^{3}$ spins on the all-to-all
($z{=}N$), the $z{=}4$ Bethe, and the square lattice
with periodic boundary conditions. The respective systems are described in detail in
Appendix \ref{AppendixMC}. For a consistent notion of ``spatial scale'' on all lattices, we
perform a graph-spectral analysis \cite{PhysRevX.13.021038}. 

Let $\mathbf{L}$ be the $N{\times} N$ symmetric Laplacian matrix of one of the
above graphs with elements $L_{ii}{=}z$, and $L_{ij}{=}-1$ when spin
$i$ and $j$ are connected and $L_{ij}{=}0$ otherwise.~The Laplacian
has $N$ orthonormal eigenvectors
$\mathbf{L}\boldsymbol{\psi}_{k}{=}l_{k}\boldsymbol{\psi}_{k},
\  k\in\{1,...,N\}$ with corresponding eigenvalues $l_{k}$ ordered 
as $l_{1}{\leq}\ldots {\leq} l_{N}$.~The lowest eigenvalue, corresponding to
$\boldsymbol{\psi}_{1}{=}N^{-1/2}(1,\ldots,1)^{\rm T}$, vanishes, i.e., $l_{1}{=}0$ 
\cite{anderson1985eigenvalues}. For the mean-field lattice, all
remaining $N{-}1$ eigenvalues are degenerate, $l_{2}{=}\ldots{=}l_{N}{=}N$
\cite{cohen2010complex}, but not for the square and Bethe lattices (see Fig.~\ref{FigA2}). 

We express the microscopic state of the lattice $\mu$ in timestep
$n\in\{0,...,n_{\rm max}\}$ as a column vector
$\boldsymbol{\sigma}^{\mu}(n)=(\sigma^{\mu}_{1}(n),...,\sigma^{\mu}_{N}(n))^{\rm T}$
and project it onto the respective eigenvectors, 
$\Psi^{\mu}_{k}(n)\equiv \boldsymbol{\psi}_{k}^{\rm T}\boldsymbol{\sigma}^{\mu}(n)$.
These spatial modes are shown as insets
in Figs.~\ref{Fig4}a-c, where we see that for the Bethe and
square lattice oscillations are pronounced on large scales (small $k$) and suppressed on small scales (large $k$). In the
mean-field system the spatial modes are equal
on all scales due to the degenerate eigenvalues.

To unravel the spatiotemporal structure, we compute the spectral
density via the discrete Fourier transform, 
\begin{equation}
    \langle |\hat{\Psi}^{\mu}_{k}(\omega) | \rangle \equiv \biggl \langle \biggl | \sum\nolimits_{n=0}^{n_{\rm max}}|\Psi^{\mu}_{k}(n)|\e{-{\rm i}2\pi \omega n/n_{\rm max}} \biggr | \biggr \rangle,
    \label{aomega}
\end{equation}
where $\langle \cdot \rangle$ indicates averaging over independent
trajectories and the absolute value takes into account that the sign of the projection is immaterial.
The results are shown in Figs.~\ref{Fig4}a-c for coherent oscillations on the $a$
lattice (those for the $b$ lattice are equivalent), with
resonances at even multiples of the respective natural frequency $\omega_0={\rm
        argmax}\langle |\hat{\Psi}^{a}_{1}(\omega)| \rangle $, which is the most dominant frequency in the spectrum and scales as $\omega_0\propto 1/\mathcal{T}$ close to the SNIPER bifurcation.~The spatiotemporal dynamics on the Bethe and square lattices is
qualitatively the same, with small-scale and high-frequency modes
suppressed. Thus, coherent oscillations are
carried by large-scale low-frequency modes, which agrees with
the large local correlations $\mathcal{C}^{\mu\mu}(t)$ shown in Fig.~\ref{Fig1}e. 
%---------------------------------------------------
%---------------------------------------------------
\section{Concluding remarks}
%---------------------------------------------------
%---------------------------------------------------
We explained the collective dynamics of the
nonreciprocal Ising system on the level of both, local and global
order beyond the mean-field approximation. A critical threshold magnitude of local order within the respective lattices was found
to control the emergence of coherent oscillations of the global order parameter. Upon increasing interactions, ghost states emerge and the residence time in either
of them eventually diverges, giving rise to a dynamically trapped
terminal state via a saddle-node-infinite-period
bifurcation. The terminal state depends on the initial condition; the
dynamics in this regime is thus non-ergodic. 

Strikingly, during coherent oscillations of global order,
a high degree of local order is preserved (see Fig.~\ref{Fig1}d). This implies nontrivial
spatiotemporal correlations between spins, confirmed by a
spectral-density maximum at large-scale low-frequency modes. In stark
contrast, on the mean-field (all-to-all) lattice there is no distinction between different spatial modes, 
annihilating any notion of spatial structure. Thus, accounting for
nearest-neighbor correlations is essential for a correct understanding of
dynamics of nonreciprocal matter with short, or more generally
finite, range of interactions.\\
\indent Our work provides a comprehensive
microscopic understanding of dynamic collective phenomena in
nonreciprocal matter without conservation laws based on the nonreciprocal Ising model. What
remains elusive are multiple ($>2$) coupled lattices,  
spatially heterogeneous/extended systems \cite{PhysRevResearch.5.013135}, as well as the
thermodynamic cost of dynamical states and bifurcations
\cite{PhysRevLett.131.258302, PhysRevX.7.021007, blom2024milestoning,
  PhysRevResearch.5.L022033}. Moreover, considering the relevance of conservation laws \cite{PhysRevLett.131.107201,PhysRevX.14.021014,PhysRevX.10.041009, GrTh2024c}, it will be essential to develop a theoretical framework for the nonreciprocal Ising model with Kawasaki dynamics. These will be addressed in future work.
%---------------------------------------------------
%---------------------------------------------------
\section{Acknowledgments}
%---------------------------------------------------
%---------------------------------------------------
Financial support from the German Research Foundation
(DFG) through the Heisenberg Program Grants GO 2762/4-1 and GO
2762/5-1 (to AG) is gratefully acknowledged. 
\appendix
%---------------------------------
%---------------------------------
%---------------------------------
\section{Derivation of Eqs.~\eqref{Eq1}-\eqref{Eq3}}\label{AppendixA}
\renewcommand{\theequation}{A\arabic{equation}}
\renewcommand{\thefigure}{A\arabic{figure}}
\setcounter{equation}{0}
\setcounter{figure}{0}
%---------------------------------
%---------------------------------
%---------------------------------
From the master equation \eqref{masterequation}, we directly obtain dynamical equations for the first two moments of single-spin values (see also Eqs.~(28) and (29) in \cite{glauber_timedependent_1963})
\begin{widetext}
\begin{align}
   \tau\frac{{\rm d}\langle \sigma^{\mu}_{i}(t) \rangle}{{\rm d}t}+\langle \sigma^{\mu}_{i}(t) \rangle  
   &=\langle \sigma^{\mu}_{i}(t)\tanh{(\Delta E^{\mu}_{i}/2)}\rangle \label{SGlaub1}, \\
   \tau\frac{{\rm d}\langle \sigma^{\mu}_{i}(t)\sigma^{\nu}_{j}(t) \rangle}{{\rm d}t}+2\langle \sigma^{\mu}_{i}(t)\sigma^{\nu}_{j}(t) \rangle
   &=\langle \sigma^{\mu}_{i}(t)\sigma^{\nu}_{j}(t)[\tanh{(\Delta E^{\mu}_{i}/2)}+\tanh{(\Delta E^{\nu}_{j}/2)}]\rangle \label{SGlaub2},
\end{align}
where $\langle f(t) \rangle \equiv
\sum_{\boldsymbol{\sigma}}P(\boldsymbol{\sigma};t)f(\boldsymbol{\sigma})$. Eqs.~\eqref{SGlaub1}-\eqref{SGlaub2}
are exact (but \emph{not} yet closed) equations for the first two moments evolving under Glauber dynamics, and will serve as our starting point to derive equations for the global and local order parameters, which we derive in two steps:
First, we sum Eqs.~\eqref{SGlaub1}-\eqref{SGlaub2} over all spins and spin pairs. Upon doing this, the left-hand side of Eqs.~\eqref{SGlaub1}-\eqref{SGlaub2} transforms into
\begin{alignat}{2}
     &\frac{1}{N}\sum_{i=1}^{N}\left(\tau\frac{{\rm d}\langle \sigma^{\mu}_{i}(t) \rangle}{{\rm d}t}+\langle \sigma^{\mu}_{i}(t) \rangle\right)  &&=\tau\frac{{\rm d} m^{\mu}(t)}{{\rm d}t}+m^{\mu}(t), \label{Sslhs1} \\
      &\frac{2}{zN}\sum_{i=1}^{N}\sum_{\langle i|j \rangle }\left(\tau\frac{{\rm d}\langle \sigma^{\mu}_{i}(t)\sigma^{\mu}_{j}(t) \rangle}{{\rm d}t}+2\langle \sigma^{\mu}_{i}(t)\sigma^{\mu}_{j}(t) \rangle\right)&&=\tau\frac{{\rm d} q^{\mu\mu}(t)}{{\rm d}t}+2q^{\mu\mu}(t),  \label{Sslhs2} \\
      &\frac{1}{N}\sum_{i=1}^{N}\left(\tau\frac{{\rm d}\langle \sigma^{a}_{i}(t)\sigma^{b}_{i}(t) \rangle}{{\rm d}t}+2\langle \sigma^{a}_{i}(t)\sigma^{b}_{i}(t) \rangle\right)&&=\tau\frac{{\rm d} q^{ab}(t)}{{\rm d}t}+2q^{ab}(t).  \label{Sslhs3}
\end{alignat}
\end{widetext}
Second, we need to evaluate the right-hand side of
Eqs.~\eqref{SGlaub1}-\eqref{SGlaub2} after summation over all spins and spin pairs. To do this, we note that $\Delta E^{\mu}_{i}$ can take on a discrete (enumerable) set of values. Consider a spin with $l\in\{0,1,...,z\}$ neighboring up spins on the same lattice and $n\in\{0,1\}$ neighboring up spins on the opposing lattice. We want to compute the change in energy upon flipping this spin. Based on Eq.~\eqref{E} we can parameterize this change in energy upon flipping the spin as
\begin{equation}
    \Delta E^{\mu}_{i}=2\sigma^{\mu}_{i} U^{\mu}_{l,n},
\end{equation}
where $U^{\mu}_{l,n}$ is given by Eq.~\eqref{U}. Using this parameterization, we evaluate the right-hand side of Eqs.~\eqref{SGlaub1}-\eqref{SGlaub2}
\begin{widetext}
\begin{alignat}{3}
    &\frac{1}{N}\sum_{i=1}^{N} \langle \sigma^{\mu}_{i}\tanh{(\Delta E^{\mu}_{i}/2)}\rangle &&= \frac{1}{N}\sum_{i=1}^{N} \langle \tanh{(U^{\mu}_{l,n})}\rangle && = \sum_{l=0}^{z}\sum_{n=0}^{1}\mathcal{P}^{\mu}_{l,n}(t) \tanh{(U^{\mu}_{l,n})}, \label{Ssrhs1}\\
    &\frac{2}{zN}\sum_{i=1}^{N}\sum_{\langle i|j \rangle} \langle \sigma^{\mu}_{i} \sigma^{\mu}_{j}\tanh{(\Delta E^{\mu}_{i}/2)}\rangle &&=  \frac{2}{zN}\sum_{i=1}^{N}\sum_{\langle i|j \rangle} \langle \sigma^{\mu}_{j}\tanh{(U^{\mu}_{l,n})}\rangle &&= \frac{2}{z}\sum_{l=0}^{z}\sum_{n=0}^{1}(2l-z)\mathcal{P}^{\mu}_{l,n}(t) \tanh{(U^{\mu}_{l,n})}, \label{Ssrhs2}\\
    &\frac{1}{N}\sum_{i=1}^{N}\langle \sigma^{a}_{i}\sigma^{b}_{i}\tanh{(\Delta E^{a}_{i}/2)}\rangle &&=  \frac{1}{N}\sum_{i=1}^{N}\langle \sigma^{b}_{i}\tanh{(U^{a}_{l,n})}\rangle &&= \sum_{l=0}^{z}\sum_{n=0}^{1}(2n-1)\mathcal{P}^{\mu}_{l,n}(t) \tanh{(U^{a}_{l,n})}.\label{Ssrhs3}
\end{alignat}
\end{widetext}
For the first equality in Eqs.~\eqref{Ssrhs1}-\eqref{Ssrhs3} we used $\tanh{(\Delta
  E^{\mu}_{i}/2)}=\tanh{(\sigma^{\mu}_{i}U^{\mu}_{l,n})}=\sigma^{\mu}_{i}\tanh{(U^{\mu}_{l,n})}$,
together with $(\sigma^{\mu}_{i})^{2}=1$. For the second equality, we
used that terms such as $\langle \tanh{(U^{\mu}_{l,n})}\rangle $
represent a weighted sum over all possible combinations of the
possible values that $\tanh{(U^{\mu}_{l,n})}$ can attain. The weights
are given by the time-dependent probability $\mathcal{P}^{\mu}_{l,n}(t)$ to find an up
or down spin with a specific local environment. By definition, this
probability is normalized $\sum_{l,n}\mathcal{P}^{\mu}_{l,n}(t)=1$. Combining Eqs.~\eqref{Sslhs1}-\eqref{Sslhs3} and \eqref{Ssrhs1}-\eqref{Ssrhs3} we obtain Eqs.~\eqref{Eq1}-\eqref{Eq3}.
%---------------------------------
%---------------------------------
\section{Derivation of Eq.~\eqref{BG3}}\label{AppendixB}
\renewcommand{\theequation}{B\arabic{equation}}
\renewcommand{\thefigure}{B\arabic{figure}}
\setcounter{equation}{0}
\setcounter{figure}{0}
%---------------------------------
%---------------------------------
Here, we derive Eq.~\eqref{BG3} based on the Bethe-Guggenheim (BG) approximation. We focus on the
probability of picking a spin on the $a$ lattice with a given specific local environment. The same reasoning will also apply for picking a spin on the $b$ lattice.  Recall that $\mathcal{P}^{a\pm}_{l,n}(t)$ is the probability at time $t$ to find an up ($+$) or down ($-$) spin with $l$ neighboring up spins on the $a$ lattice and $n$ neighboring up spins on the $b$ lattice. On the BG level, we assume ideal mixing of nearest-neighbor spin pairs, resulting in the following expressions
\begin{widetext}
\begin{align}
    \mathcal{P}^{a+}_{l,n}&=\underbrace{[N^{a}_{+}/N]}_{\substack{\rm probability \\ \substack{\rm for \ up \ spin \\ {\rm on \ the} \ a \ {\rm lattice}}}}  \times 
    \underbrace{\left[\binom{N^{aa}_{++}}{l}\binom{N^{aa}_{+-}/2}{z-l}/\binom{N^{aa}_{++}+N^{aa}_{+-}/2}{z}\right]}_{\substack{\rm probability \ for \\   \substack{l \ {\rm \  neighboring \ up} \\ {\rm spins \ on \ the} \ a \ {\rm lattice}}}} \times  \underbrace{\left[\binom{N^{ab}_{++}}{n}\binom{N^{ab}_{+-}}{1-n}/\binom{N^{ab}_{++}+N^{ab}_{+-}}{1}\right]}_{\substack{\rm probability \ for \\   \substack{n \ {\rm \  neighboring \ up} \\ {\rm spins \ on \ the } \ b \ {\rm lattice}}}}, \label{SBG0} \\ \nonumber \\
    \mathcal{P}^{a-}_{l,n}&=\underbrace{[N^{a}_{-}/N]}_{\substack{\rm probability \\ \substack{\rm for \ down \ spin \\ {\rm on \ the} \ a \ {\rm lattice}}}}  \times 
    \underbrace{\left[\binom{N^{aa}_{-+}/2}{l}\binom{N^{aa}_{--}}{z-l}/\binom{N^{aa}_{-+}/2+N^{aa}_{--}}{z}\right]}_{\substack{\rm probability \ for \\   \substack{l \ {\rm \  neighboring \ up} \\ {\rm spins \ on \ the} \ a \ {\rm lattice}}}} \times  \underbrace{\left[\binom{N^{ab}_{-+}}{n}\binom{N^{ab}_{--}}{1-n}/\binom{N^{ab}_{-+}+N^{ab}_{--}}{1}\right]}_{\substack{\rm probability \ for \\   \substack{n \ {\rm \  neighboring \ up} \\ {\rm spins \ on \ the} \ b \ {\rm lattice}}}},
    \label{SBG1}
\end{align}
\end{widetext}
where, for example, $N^{ab}_{+-}$ is the total number of nearest-neighbor spin pairs with an up spin on the $a$ lattice and a down spin on the $b$ lattice. To relate $N^{ab}_{+-}$ and the other spin pair numbers to the global and local order, we make use of the following exact relations for periodic lattices
\begin{align}
    2N^{\mu\mu}_{\pm\pm}+N^{\mu\mu}_{+-}&=zN^{\mu}_{\pm}, \label{Srel1} \\
    N^{ab}_{\pm \pm }+N^{ab}_{\pm \mp}&=N^{a}_{\pm }, 
\end{align}
in combination with
\begin{align}
    N^{\mu}_{\pm}&=N(1\pm m^{\mu})/2. \label{SrelN}
\end{align}
Furthermore, we use the definition of local order given by Eqs.~\eqref{qmumu}-\eqref{qmunu} to write
\begin{align}
    q^{\mu\mu}&=2(N^{\mu\mu}_{++}+N^{\mu\mu}_{--}-N^{\mu\mu}_{+-})/zN\nonumber \\
    &=1-4N^{\mu\mu}_{+-}/zN,
    \\
    q^{ab}&=(N^{ab}_{++}+N^{ab}_{--}-N^{ab}_{+-}-N^{ab}_{-+})/N\nonumber \\
    &=1-2N^{ab}_{+-}/N-2N^{ab}_{-+}/N \nonumber \\
    &=1+m^{a}-m^{b}-4N^{ab}_{+-}/N, \label{Srelf}
\end{align}
where in the last line we used the relation
\begin{equation}
    m^{b}-m^{a}=2(N^{ab}_{-+}-N^{ab}_{+-})/N.
\end{equation}
Using the relations \eqref{Srel1}-\eqref{Srelf} we obtain the following expression for the spin pairs within the same lattice
\begin{align}
N^{\mu\mu}_{\pm\pm}
&=(z/8)N(1\pm2m^{\mu}+q^{\mu\mu}), \label{Nmumu}  \\
N^{\mu\mu}_{\pm \mp}&=(z/4)N(1-q^{\mu\mu}), \
\end{align}
and for the spin pairs between the two opposing lattices
\begin{align}
    N^{ab}_{\pm\pm}
    &=(1/4)N(1\pm m^{a}\pm m^{b}+q^{ab}), \\
        N^{ab}_{\pm\mp}
    &=(1/4)N(1\pm m^{a}\mp m^{b}-q^{ab}). \label{Nab} 
\end{align}
Inserting Eqs.~\eqref{Nmumu}-\eqref{Nab} into Eqs.~\eqref{SBG0}-\eqref{SBG1} and
taking the thermodynamic limit $N\rightarrow \infty$ while keeping
$m^{\mu}(t)$, $q^{\mu\mu}(t)$, and $q^{ab}(t)$ fixed, we obtain Eq.~\eqref{BG3}.
%---------------------------------
%---------------------------------
\section{Steady-state local order}\label{AppendixQ}
\renewcommand{\theequation}{C\arabic{equation}}
\renewcommand{\thefigure}{C\arabic{figure}}
\setcounter{equation}{0}
\setcounter{figure}{0}
%---------------------------------
%---------------------------------
The trivial steady state is given by the disordered state with
$m^{\mu}_{s}=0$ and $q^{ab}_{s}=0$. To solve for the steady state of the local order, denoted as $q_{s}(J,K)$, we need to solve
\begin{equation}
    q_{s}(J,K)=\frac{1}{z}\sum_{l=0}^{z}\sum_{n=0}^{1}(2l-z)(\overline{\mathcal{P}}^{+}_{l}+\overline{\mathcal{P}}^{-}_{l})\tanh{(U^{a}_{l,n})}, \label{Sqbg} 
\end{equation}
where $\overline{\mathcal{P}}^{\pm}_{l}(q_s)$ are the probabilities \eqref{BG3} evaluated at steady-state values given by Eq.~\eqref{overlineP}. Equation \eqref{Sqbg} can be solved for specific integer values of $z$. For example, for $z=2$ we obtain
\begin{equation*}
    q_{s}(J,K)|_{z=2}=\frac{2-\sqrt{4-\left[\sum_{n=\pm}\tanh{(2J+nK)}\right]^{2}}}{\sum_{n=\pm}\tanh{(2J+nK)}}.
\end{equation*}
For $z=4$, the solution can be written as
\begin{widetext}
\begin{equation}
     q_{s}(J,K)|_{z=4}=\mathcal{S}(J,K)-(1/2)\sqrt{-4\mathcal{S}(J,K)^{2}+2\mathcal{H}(J,K)+\mathcal{Q}(J,K)/\mathcal{S}(J,K)},
\end{equation}
where we have introduced the auxiliary functions
\begin{align}
    \mathcal{H}(J,K)&\equiv \frac{3\cosh{(4J)}[\cosh{(4J)}+\cosh{(2K)}]}{\sinh^{2}{(2J)}[\cosh{(4J)}-2\sinh^{2}{(K)}]}, \\
    \mathcal{Q}(J,K)&\equiv 16\left(\sum\nolimits_{n=\pm}[\tanh{(4J+nK)}-2\tanh{(2J+nK)}]\right)^{-1}, \\
    \mathcal{S}(J,K)&\equiv (1/2)\sqrt{(2/3)\mathcal{H}(J,K)+(\mathcal{Q}(J,K)/6)\left(\mathcal{A}(J,K)+\Delta_{0}(J,K)/\mathcal{A}(J,K)\right)},\\
    \mathcal{A}(J,K)&\equiv2^{-1/3}\left(\Delta_{1}(J,K)+\sqrt{\Delta^{2}_{1}(J,K)-4\Delta^{3}_{0}(J,K)}\right)^{1/3}, \\
    \Delta_{0}(J,K)&\equiv -(3/4)\sum\nolimits_{n=\pm}[\tanh{(2J{+}nK)}+\tanh{(4J{+}nK)}]\sum\nolimits_{n=\pm}[\tanh{(2J{+}nK)}-\tanh{(4J{+}nK)}], \\
    \Delta_{1}(J,K)&\equiv \! 216\cosh{(2J)}\!\cosh{(2K)}\!\sech{(4J{-}K)}\!\sech{(4J{+}K)}\!\sinh^{3}{(2J)}\sinh^{2}{(K)}[\cosh{(4J)}{+}\cosh{(2K)}]^{-2}.
\end{align}
%---------------------------------
%---------------------------------
\section{Oscillatory Instability}\label{AppendixD}
\renewcommand{\theequation}{D\arabic{equation}}
\renewcommand{\thefigure}{D\arabic{figure}}
\setcounter{equation}{0}
\setcounter{figure}{0}
%---------------------------------
%---------------------------------
Here, we prove that $M_{2}(q_s;J,K\neq
0)\neq 0$. To see this, we explicitly write out the first sum over
$n$ in Eq.~\eqref{d1}
\begin{equation}
    M_{2}(q_s;J,K)=
    \sum_{l=0}^{z}(\overline{\mathcal{P}}^{+}_{l}+\overline{\mathcal{P}}^{-}_{l})\left(\tanh{([2l{-}z]J{+}K)}{-}\tanh{([2l{-}z]J{-}K)}\right).
\end{equation}
\end{widetext}
Note that $\overline{\mathcal{P}}^{\pm}_{l}(q_s)> 0$ for $q_{s}\in(-1,1)$, which follows straightforwardly from Eq.~\eqref{overlineP}. Furthermore, since $\tanh{(x)}$ is an increasing function of $x$, we have $\tanh{([2l-z]J+K)}-\tanh{([2l-z]J-K)}> 0$ for $ K>0$ and $\tanh{([2l-z]J+K)}-\tanh{([2l-z]J-K)}< 0$ for $ K<0$. Hence, $M_{2}(q_s;J,K)$
is given by a sum over strictly positive (for $K>0$) or negative (for
$K<0$) terms, rendering $M_{2}(q_s;J,K\neq0)\neq 0$. This
results in complex eigenvalues for $\lambda_{\pm}(q_s;J,K)$ as shown in Fig.~\ref{Fig1}b.
%---------------------------------
%---------------------------------
\section{Mean-field approximation}\label{AppendixMF}
\renewcommand{\theequation}{E\arabic{equation}}
\renewcommand{\thefigure}{E\arabic{figure}}
\setcounter{equation}{0}
\setcounter{figure}{0}
%---------------------------------
%---------------------------------
A less accurate
technique to obtain approximate evolution equations 
is the mean-field (MF) approximation (originally developed in \cite{Penrose1991}), where one makes the rudimentary (uncontrolled) assumption
\begin{equation}
    \langle \tanh{\left(\Delta E^{\mu}_{i}/2\right)} \rangle \approx \tanh{\langle \Delta E^{\mu}_{i}/2 \rangle},
    \label{MFassumption}
\end{equation}
yielding the evolution equations 
\begin{align}
    \tau \frac{{\rm d}m^{a}(t)}{{\rm d}t}+m^{a}(t)&=\tanh{(zJ_{a} m^{a}(t)+K_{a}m^{b}(t))}, \nonumber \\
    \tau \frac{{\rm d}m^{b}(t)}{{\rm d}t}+m^{b}(t)&=\tanh{(zJ_{b} m^{b}(t)+K_{b}m^{a}(t))},
    \label{EqMF}
\end{align}
which are exact on the fully connected mean-field lattice, where the
local order is trivial (i.e.\ there is \emph{no} sense of ``local''),
$q^{\mu\nu}(t)=m^{\mu}(t)m^{\nu}(t)$, and therefore
$\mathcal{C}^{\mu\nu}(t)=0$. A linear stability analysis around the
trivial steady state $m^{\mu}_{s}=0$ for $J_{a}=J_{b}=J$ and
$K_{a}=-K_{b}=K$ leads to a linear stability equation where the eigenvalues of the linear stability matrix are given by
\begin{equation}
    \lambda^{\rm MF}_{\pm}(J,K)=(zJ-1)\pm iK.
\end{equation}
Hence, the Hopf bifurcation occurs at $J=1/z$ and $K\neq 0$, such that
${\rm Re}(\lambda^{\rm MF}_{\pm})=0$ and ${\rm Im}(\lambda^{\rm
  MF}_{\pm})\neq 0$. This corresponds to a straight vertical line in
the $(J,K)$-plane, as shown in Fig.~\ref{FigA1}a and also found in \cite{avni2023nonreciprocal}.~Notably, in the MF approximation we do \emph{not} observe a critical value for local order, which is present in the more accurate BG approximation.
\begin{figure}
    \centering
    \includegraphics[width=\linewidth]{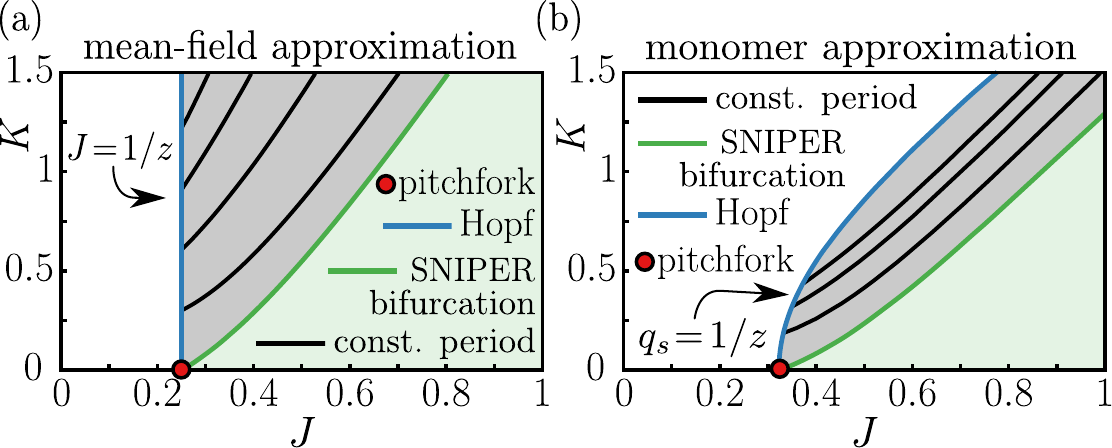}
    \caption{Phase diagram for global order $m^{\mu}_{s}$ obtained
      with the mean-field (MF) approximation (a) and the monomer
      approximation (b) for the perfect nonreciprocal setting with
      $J_{a}=J_{b}=J$ and $K_{a}=-K_{b}=K$. In the MF approximation
      the Hopf bifurcation (blue line) is set by $J=1/z$. In the
      monomer approximation, the Hopf bifurcations (blue line) is set
      by a critical local order $q_{s}=1/z$, which is more similar to the Bethe-Guggenheim approximation where $q_{s}=1/(z-1)$ (see Fig.~\ref{Fig2}d).}
    \label{FigA1}
\end{figure}\\
\begin{table}
   \caption{Overview of approximation techniques and their dynamical equations}
        \centering
        \begin{tabular}{|p{1.4cm}||p{1.3cm}|p{2.5cm}|p{1.7cm}|}
            \multicolumn{4}{}{} \\
            \hline
            approx. technique & $m^{\mu}(t)$ & $q^{\mu\nu}(t)$ &  Hopf \\
            \hline 
            MF & \eqref{EqMF} & $=m^{\mu}(t)m^{\nu}(t)$ & $J{=}1/z$ \\
            monomer &  \eqref{Eq1}+\eqref{MMP} & \eqref{Eq2}-\eqref{Eq3}+\eqref{MMP}; slaved by $m^{\mu}(t)$ & $q_{s}{=}1/z$ \\
            BG & \eqref{Eq1}+\eqref{BG3} & \eqref{Eq2}-\eqref{Eq3}+\eqref{BG3}; not slaved & $q_{s}{=}1/(z{-}1)$ \\
            \hline
        \end{tabular}
        \label{Table I}
\end{table}
%---------------------------------
%---------------------------------
\section{Monomer approximation}\label{AppendixMon}
\renewcommand{\theequation}{F\arabic{equation}}
\renewcommand{\thefigure}{F\arabic{figure}}
\setcounter{equation}{0}
\setcounter{figure}{0}
%---------------------------------
%---------------------------------
Another approximation
technique we developed in this work is what we call the ``monomer
approximation''.~It is more accurate than the MF but less accurate
than the BG approximation.  
 
The conceptual difference between the MF on the one hand, and the monomer and BG
approximations on the other hand, lies in the treatment of the average
$\langle \tanh{\left(\Delta E^{\mu}_{i}/2\right)} \rangle $. Whereas
the MF approximation simply moves the average to the argument as shown
in Eq.~\eqref{MFassumption}, the monomer and BG approximations use the fact that the value of $\Delta E^{\mu}_{i}/2$ lies in an enumerable set given by $U^{\mu}_{l,n}\equiv[2l-z]J_{\mu}+[2n-1]K_{\mu}$ with $l\in\{0,..,z\}$ and $n\in\{0,1\}$. This allows for an explicit summation
\begin{equation}
    \langle \tanh{\left(\Delta E^{\mu}_{i}/2\right)} \rangle =\sum_{l=0}^{z}\sum_{n=0}^{1}\mathcal{P}^{\mu}_{l,n}(t)\tanh{(U^{\mu}_{l,n})},
\end{equation}
where only the probability $\mathcal{P}^{\mu}_{l,n}(t)$ has to approximated. 
\renewcommand{\thefigure}{G\arabic{figure}}
\begin{figure*}[t!]
    \centering
    \includegraphics[width=\textwidth]{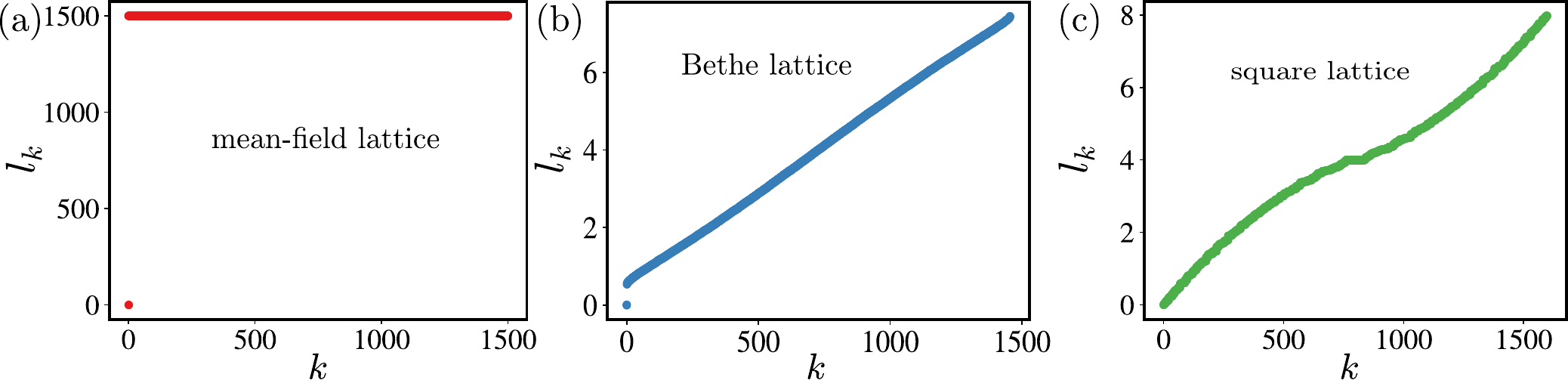}
    \caption{Eigenvalues $l_{k}$ of the Laplacian matrix $\mathbf{L}$
      for the mean-field lattice (a), Bethe lattice (b), and square
      lattice (c). For the mean-field lattice, all eigenvalues except
      for the first are degenerate with value $l_{2}=...=l_{N}=N$,
      where $N$ is the system size given in table \ref{STable I}. The
      Bethe lattice has a spectral gap between the lowest eigenvalue $l_{1}=0$ and $l_{2}$.}
    \label{FigA2}
\end{figure*}

Similar to the BG approximation, the resulting evolution equations in the monomer approximation are governed by Eqs.~\eqref{Eq1}-\eqref{Eq3}, but the time-dependent probabilities are different and read
(the derivation is given in \cite{Note2})
\begin{equation}
     \mathcal{P}^{a }_{l,n}(t)
    =\frac{2\mathcal{C}^{z}_{l}(1+m^{a}(t))^{l}(1+m^{b}(t))^{n}}{(1-m^{a}(t))^{l-z}(1-m^{b}(t))^{n-1}},
    \label{MMP}
\end{equation}
and $ \mathcal{P}^{b}_{l,n}(t)$ is obtained by replacing $m^{a}(t)$ with $m^{b}(t)$ in Eq.~\eqref{MMP}. Since $\mathcal{P}^{a}_{l,n}(t)$ is independent of the local order $q^{\mu\nu}(t)$, this implies that $q^{\mu\nu}(t)$ is slaved by $m^{\mu}(t)$, however, $q^{\mu\nu}(t)\neq m^{\mu}(t)m^{\nu}(t)$. Performing a linear stability analysis around the trivial steady state $m^{\mu}_{s}=0$ for $J_{a}=J_{b}=J$ and $K_{a}=-K_{b}=K$, we obtain a linear stability equation where the eigenvalues of the linear stability matrix 
can neatly be written as
\begin{equation}
    \hat{\lambda}_{\pm}(q_s;J,K)=(zq_{s}(J,K)-1)\pm i\hat{M}_{2}(J,K),\end{equation}
where $q_{s}(J,K)$ is the steady-state value of the local order in the monomer approximation, and
\begin{equation}
    \hat{M}_{2}(J,K)=2\sum_{l=0}^{z}\sum_{n=0}^{1}(2n-1)\mathcal{C}^{z}_{l}\tanh{(U^{a}_{l,n})}.
\end{equation}
From this follows that the Hopf bifurcation occurs at
$q_{s}=1/z$ and $K\neq 0$, such that ${\rm
  Re}(\hat{\lambda}_{\pm})=0$ and ${\rm Im}(\hat{\lambda}_{\pm})\neq
0$. Hence, as within the BG approximation, we also find the existence
of a critical local order in the monomer approximation. Contrary
to the MF approximation, the Hopf line is \emph{not} a straight vertical line in the $(J,K)$-plane, as shown in Fig.~\ref{FigA1}b.

To provide a concise overview of the various approximation
techniques, we summarize in Table~\ref{Table I} the respective
evolution and conditions for the Hopf bifurcation. 
%---------------------------------
%---------------------------------
%---------------------------------
\section{Kinetic Monte-Carlo simulations}\label{AppendixMC}
\renewcommand{\theequation}{G\arabic{equation}}
\renewcommand{\thesubsection}{G\arabic{subsection}}
\setcounter{equation}{0}
\setcounter{figure}{0}
%---------------------------------
%---------------------------------
%---------------------------------
For the results shown in Fig.~\ref{Fig1}c,d (black dashed lines) and Fig.~\ref{Fig4} we performed kinetic MC simulations on three different types of
lattices: (i) the fully connected mean-field (MF) lattice, (ii) the Bethe lattice, and (iii) the square lattice with periodic boundary conditions. Simulations on the Bethe lattice were performed using the random graph algorithm \cite{D_A_Johnston_1998SM, Deepak_Dhar_1997SM}, which works as follows: Consider a Bethe lattice with coordination number $z$. First, we create a Cayley tree of $i=\{1,...,N\}$ spins with coordination number $z$. The spins on the outer layer are connected to one spin on the inner layer. To create the remaining $z-1$ connections, we randomly pair spins on the outer layer to other spins on the outer layer. The final result is a Cayley tree with random connections on the outer layer. Note that for both lattices $a$ and $b$ we create new random connections. For large $N$, it has been shown that the Ising model on an ensemble of such random graphs is equivalent to the Ising model on a Bethe lattice \cite{D_A_Johnston_1998SM}. Indeed, for large $N$ we find perfect agreement between the simulations and our theory, as shown in Fig.~\ref{Fig1}c,d. For the MF lattice, we connect all spins with each other, resulting in a fully connected graph.
%---------------------------------
%---------------------------------
\subsection{Simulation setup}
%---------------------------------
%---------------------------------
In Table~\ref{STable I} we summarize the size of the system, the number of
trajectories, and the parameter settings that were used to obtain the
spectral density shown in Fig.~\ref{Fig4}. As initial conditions,
we selected a randomly mixed configuration of up- and down-spins for
fixed magnetization.
\begin{table}
   \caption{Simulation parameters for results shown in Fig.~\ref{Fig4}.}
        \centering
        \begin{tabular}{|p{1.6cm}||p{1.2cm}|p{1.4cm}|p{1.1cm}|p{1.2cm}|p{1.2cm}|}
            \multicolumn{6}{}{} \\
            \hline
            lattice & size ($N$) & MC steps & \# traj. & $J$ $[k_{\rm B}T]$ & $K$ $[k_{\rm B}T]$  \\
            \hline 
            mean-field & $1500$ & $N \times 10^{3}$ & $500$  & $1.5/N$ & $0.3$\\
            Bethe & $1457$ & $N\times 10^{3}$ & $500$  & $0.5$ & $0.3$\\
            square & $40\times40$ & $N\times 10^{3}$ & $500$ & $0.6$ & $0.3$\\
            \hline
        \end{tabular}
        \label{STable I}
\end{table}

For the results shown in Fig.~\ref{Fig1}c,d we used a Bethe lattice with
system size $N=118097$. Such a large system size was not feasible for
the setup of
Fig.~\ref{Fig4} since the spectral density $\langle
|\hat{\Psi}^{\mu}_{k}(\omega) | \rangle$ must be averaged over many independent trajectories, resulting in memory issues for too large $N$.
%---------------------------------
%---------------------------------
\subsection{Eigenvalues of Laplacian matrix}
%---------------------------------
%---------------------------------
In Fig.~\ref{Fig4} we plot the spectral density $\langle |\hat{\Psi}^{\mu}_{k}(\omega) | \rangle$ as a function of the eigenvalues $l_{k}$ of the Laplacian matrix $\mathbf{L}$. To obtain the eigenvalues, we numerically diagonalized the Laplacian $\mathbf{L}$ in Python, and the resulting eigenvalues are shown in Fig.~\ref{FigA2}. Note that for the mean-field lattices, all eigenvalues except the first are degenerate and equal to $l_{2}=...=l_{N}=N$.
 %---------------------------------
%---------------------------------
\subsection{Animations}
%---------------------------------
%---------------------------------
To visualize the dynamics of the nonreciprocal Ising model on the mean-field lattice, Bethe lattice, and square lattice, we have provided animations in \cite{Note2}. In each animation, we show three independent simulations on the aforementioned lattices, with the coupling strengths $(J,K)$ reported in Table \ref{STable I}. More information about the animations is given below:
\begin{itemize}
    \item “MF\_lattice.gif” shows simulations on the mean-field lattice for $N=2000$ spins, where each spin is connected to every other spin. For illustrative purposes, the edges between spins are not shown. 
    \item “Bethe\_lattice.gif” shows simulations on the Bethe lattice for $N=13121$ spins, corresponding to a Bethe lattice with $8$ layers and a coordination number of $z=4$. For illustrative purposes, only the first $6$ layers of the Bethe lattice are shown. 
    \item “Square\_lattice.gif” shows simulations on the square lattice for $N=122 \times 122$ spins. 
\end{itemize}
\setcounter{tocdepth}{1}
%---------------------------------
%---------------------------------
%---------------------------------
%- --------------------------------
\let\oldaddcontentsline\addcontentsline
\renewcommand{\addcontentsline}[3]{}
\bibliographystyle{apsrev4-2.bst}
\bibliography{nonreciprocal.bib}

\begin{thebibliography}{48}%
\makeatletter
\providecommand \@ifxundefined [1]{%
 \@ifx{#1\undefined}
}%
\providecommand \@ifnum [1]{%
 \ifnum #1\expandafter \@firstoftwo
 \else \expandafter \@secondoftwo
 \fi
}%
\providecommand \@ifx [1]{%
 \ifx #1\expandafter \@firstoftwo
 \else \expandafter \@secondoftwo
 \fi
}%
\providecommand \natexlab [1]{#1}%
\providecommand \enquote  [1]{``#1''}%
\providecommand \bibnamefont  [1]{#1}%
\providecommand \bibfnamefont [1]{#1}%
\providecommand \citenamefont [1]{#1}%
\providecommand \href@noop [0]{\@secondoftwo}%
\providecommand \href [0]{\begingroup \@sanitize@url \@href}%
\providecommand \@href[1]{\@@startlink{#1}\@@href}%
\providecommand \@@href[1]{\endgroup#1\@@endlink}%
\providecommand \@sanitize@url [0]{\catcode `\\12\catcode `\$12\catcode `\&12\catcode `\#12\catcode `\^12\catcode `\_12\catcode `\%12\relax}%
\providecommand \@@startlink[1]{}%
\providecommand \@@endlink[0]{}%
\providecommand \url  [0]{\begingroup\@sanitize@url \@url }%
\providecommand \@url [1]{\endgroup\@href {#1}{\urlprefix }}%
\providecommand \urlprefix  [0]{URL }%
\providecommand \Eprint [0]{\href }%
\providecommand \doibase [0]{https://doi.org/}%
\providecommand \selectlanguage [0]{\@gobble}%
\providecommand \bibinfo  [0]{\@secondoftwo}%
\providecommand \bibfield  [0]{\@secondoftwo}%
\providecommand \translation [1]{[#1]}%
\providecommand \BibitemOpen [0]{}%
\providecommand \bibitemStop [0]{}%
\providecommand \bibitemNoStop [0]{.\EOS\space}%
\providecommand \EOS [0]{\spacefactor3000\relax}%
\providecommand \BibitemShut  [1]{\csname bibitem#1\endcsname}%
\let\auto@bib@innerbib\@empty
%</preamble>
\bibitem [{\citenamefont {Guislain}\ and\ \citenamefont {Bertin}(2024{\natexlab{a}})}]{Guislain_2024A}%
  \BibitemOpen
  \bibfield  {author} {\bibinfo {author} {\bibfnamefont {L.}~\bibnamefont {Guislain}}\ and\ \bibinfo {author} {\bibfnamefont {E.}~\bibnamefont {Bertin}},\ }\href {https://doi.org/10.1088/1742-5468/ad72dc} {\bibfield  {journal} {\bibinfo  {journal} {J. Stat. Mech.: Theory Exp.}\ }\textbf {\bibinfo {volume} {2024}}\bibinfo  {number} { (9)},\ \bibinfo {pages} {093210}}\BibitemShut {NoStop}%
\bibitem [{\citenamefont {Guislain}\ and\ \citenamefont {Bertin}(2024{\natexlab{b}})}]{Guislain_2024B}%
  \BibitemOpen
\bibfield  {number} {  }\bibfield  {author} {\bibinfo {author} {\bibfnamefont {L.}~\bibnamefont {Guislain}}\ and\ \bibinfo {author} {\bibfnamefont {E.}~\bibnamefont {Bertin}},\ }\href {https://doi.org/10.1088/1751-8121/ad6ab4} {\bibfield  {journal} {\bibinfo  {journal} {J. Phys. A}\ }\textbf {\bibinfo {volume} {57}},\ \bibinfo {pages} {375001} (\bibinfo {year} {2024}{\natexlab{b}})}\BibitemShut {NoStop}%
\bibitem [{\citenamefont {Seara}\ \emph {et~al.}(2023)\citenamefont {Seara}, \citenamefont {Piya},\ and\ \citenamefont {Tabatabai}}]{Seara_2023}%
  \BibitemOpen
  \bibfield  {author} {\bibinfo {author} {\bibfnamefont {D.~S.}\ \bibnamefont {Seara}}, \bibinfo {author} {\bibfnamefont {A.}~\bibnamefont {Piya}},\ and\ \bibinfo {author} {\bibfnamefont {A.~P.}\ \bibnamefont {Tabatabai}},\ }\href {https://doi.org/10.1088/1742-5468/accce7} {\bibfield  {journal} {\bibinfo  {journal} {J. Stat. Mech.: Theory Exp.}\ }\textbf {\bibinfo {volume} {2023}}\bibinfo  {number} { (4)},\ \bibinfo {pages} {043209}}\BibitemShut {NoStop}%
\bibitem [{\citenamefont {Collet}\ \emph {et~al.}(2016)\citenamefont {Collet}, \citenamefont {Formentin},\ and\ \citenamefont {Tovazzi}}]{PhysRevE.94.042139}%
  \BibitemOpen
\bibfield  {number} {  }\bibfield  {author} {\bibinfo {author} {\bibfnamefont {F.}~\bibnamefont {Collet}}, \bibinfo {author} {\bibfnamefont {M.}~\bibnamefont {Formentin}},\ and\ \bibinfo {author} {\bibfnamefont {D.}~\bibnamefont {Tovazzi}},\ }\href {https://doi.org/10.1103/PhysRevE.94.042139} {\bibfield  {journal} {\bibinfo  {journal} {Phys. Rev. E}\ }\textbf {\bibinfo {volume} {94}},\ \bibinfo {pages} {042139} (\bibinfo {year} {2016})}\BibitemShut {NoStop}%
\bibitem [{\citenamefont {Collet}(2014)}]{Collet2014}%
  \BibitemOpen
  \bibfield  {author} {\bibinfo {author} {\bibfnamefont {F.}~\bibnamefont {Collet}},\ }\href {https://doi.org/10.1007/s10955-014-1105-9} {\bibfield  {journal} {\bibinfo  {journal} {J. Stat. Phys.}\ }\textbf {\bibinfo {volume} {157}},\ \bibinfo {pages} {1301–1319} (\bibinfo {year} {2014})}\BibitemShut {NoStop}%
\bibitem [{\citenamefont {Collet}\ and\ \citenamefont {Formentin}(2019)}]{Collet2019}%
  \BibitemOpen
  \bibfield  {author} {\bibinfo {author} {\bibfnamefont {F.}~\bibnamefont {Collet}}\ and\ \bibinfo {author} {\bibfnamefont {M.}~\bibnamefont {Formentin}},\ }\href {https://doi.org/10.1007/s10955-019-02310-7} {\bibfield  {journal} {\bibinfo  {journal} {J. Stat. Phys.}\ }\textbf {\bibinfo {volume} {176}},\ \bibinfo {pages} {478–491} (\bibinfo {year} {2019})}\BibitemShut {NoStop}%
\bibitem [{\citenamefont {Avni}\ \emph {et~al.}(2023)\citenamefont {Avni}, \citenamefont {Fruchart}, \citenamefont {Martin}, \citenamefont {Seara},\ and\ \citenamefont {Vitelli}}]{avni2023nonreciprocal}%
  \BibitemOpen
  \bibfield  {author} {\bibinfo {author} {\bibfnamefont {Y.}~\bibnamefont {Avni}}, \bibinfo {author} {\bibfnamefont {M.}~\bibnamefont {Fruchart}}, \bibinfo {author} {\bibfnamefont {D.}~\bibnamefont {Martin}}, \bibinfo {author} {\bibfnamefont {D.}~\bibnamefont {Seara}},\ and\ \bibinfo {author} {\bibfnamefont {V.}~\bibnamefont {Vitelli}},\ }\href@noop {} {\bibinfo {title} {The non-reciprocal {I}sing model}} (\bibinfo {year} {2023}),\ \Eprint {https://arxiv.org/abs/2311.05471} {arXiv:2311.05471 [cond-mat.stat-mech]} \BibitemShut {NoStop}%
\bibitem [{\citenamefont {Loos}\ \emph {et~al.}(2023)\citenamefont {Loos}, \citenamefont {Klapp},\ and\ \citenamefont {Martynec}}]{PhysRevLett.130.198301}%
  \BibitemOpen
  \bibfield  {author} {\bibinfo {author} {\bibfnamefont {S.~A.~M.}\ \bibnamefont {Loos}}, \bibinfo {author} {\bibfnamefont {S.~H.~L.}\ \bibnamefont {Klapp}},\ and\ \bibinfo {author} {\bibfnamefont {T.}~\bibnamefont {Martynec}},\ }\href {https://doi.org/10.1103/PhysRevLett.130.198301} {\bibfield  {journal} {\bibinfo  {journal} {Phys. Rev. Lett.}\ }\textbf {\bibinfo {volume} {130}},\ \bibinfo {pages} {198301} (\bibinfo {year} {2023})}\BibitemShut {NoStop}%
\bibitem [{\citenamefont {Guislain}\ and\ \citenamefont {Bertin}(2024{\natexlab{c}})}]{PhysRevE.109.034131}%
  \BibitemOpen
  \bibfield  {author} {\bibinfo {author} {\bibfnamefont {L.}~\bibnamefont {Guislain}}\ and\ \bibinfo {author} {\bibfnamefont {E.}~\bibnamefont {Bertin}},\ }\href {https://doi.org/10.1103/PhysRevE.109.034131} {\bibfield  {journal} {\bibinfo  {journal} {Phys. Rev. E}\ }\textbf {\bibinfo {volume} {109}},\ \bibinfo {pages} {034131} (\bibinfo {year} {2024}{\natexlab{c}})}\BibitemShut {NoStop}%
\bibitem [{\citenamefont {Osat}\ and\ \citenamefont {Golestanian}(2023)}]{osat2023non}%
  \BibitemOpen
  \bibfield  {author} {\bibinfo {author} {\bibfnamefont {S.}~\bibnamefont {Osat}}\ and\ \bibinfo {author} {\bibfnamefont {R.}~\bibnamefont {Golestanian}},\ }\href {https://doi.org/10.1038/s41565-022-01258-2} {\bibfield  {journal} {\bibinfo  {journal} {Nat. Nanotechnol.}\ }\textbf {\bibinfo {volume} {18}},\ \bibinfo {pages} {79} (\bibinfo {year} {2023})}\BibitemShut {NoStop}%
\bibitem [{\citenamefont {Osat}\ \emph {et~al.}(2024)\citenamefont {Osat}, \citenamefont {Metson}, \citenamefont {Kardar},\ and\ \citenamefont {Golestanian}}]{PhysRevLett.133.028301}%
  \BibitemOpen
  \bibfield  {author} {\bibinfo {author} {\bibfnamefont {S.}~\bibnamefont {Osat}}, \bibinfo {author} {\bibfnamefont {J.}~\bibnamefont {Metson}}, \bibinfo {author} {\bibfnamefont {M.}~\bibnamefont {Kardar}},\ and\ \bibinfo {author} {\bibfnamefont {R.}~\bibnamefont {Golestanian}},\ }\href {https://doi.org/10.1103/PhysRevLett.133.028301} {\bibfield  {journal} {\bibinfo  {journal} {Phys. Rev. Lett.}\ }\textbf {\bibinfo {volume} {133}},\ \bibinfo {pages} {028301} (\bibinfo {year} {2024})}\BibitemShut {NoStop}%
\bibitem [{\citenamefont {Ivlev}\ \emph {et~al.}(2015)\citenamefont {Ivlev}, \citenamefont {Bartnick}, \citenamefont {Heinen}, \citenamefont {Du}, \citenamefont {Nosenko},\ and\ \citenamefont {L\"owen}}]{PhysRevX.5.011035}%
  \BibitemOpen
  \bibfield  {author} {\bibinfo {author} {\bibfnamefont {A.~V.}\ \bibnamefont {Ivlev}}, \bibinfo {author} {\bibfnamefont {J.}~\bibnamefont {Bartnick}}, \bibinfo {author} {\bibfnamefont {M.}~\bibnamefont {Heinen}}, \bibinfo {author} {\bibfnamefont {C.-R.}\ \bibnamefont {Du}}, \bibinfo {author} {\bibfnamefont {V.}~\bibnamefont {Nosenko}},\ and\ \bibinfo {author} {\bibfnamefont {H.}~\bibnamefont {L\"owen}},\ }\href {https://doi.org/10.1103/PhysRevX.5.011035} {\bibfield  {journal} {\bibinfo  {journal} {Phys. Rev. X}\ }\textbf {\bibinfo {volume} {5}},\ \bibinfo {pages} {011035} (\bibinfo {year} {2015})}\BibitemShut {NoStop}%
\bibitem [{\citenamefont {Brauns}\ and\ \citenamefont {Marchetti}(2024)}]{PhysRevX.14.021014}%
  \BibitemOpen
  \bibfield  {author} {\bibinfo {author} {\bibfnamefont {F.}~\bibnamefont {Brauns}}\ and\ \bibinfo {author} {\bibfnamefont {M.~C.}\ \bibnamefont {Marchetti}},\ }\href {https://doi.org/10.1103/PhysRevX.14.021014} {\bibfield  {journal} {\bibinfo  {journal} {Phys. Rev. X}\ }\textbf {\bibinfo {volume} {14}},\ \bibinfo {pages} {021014} (\bibinfo {year} {2024})}\BibitemShut {NoStop}%
\bibitem [{\citenamefont {te~Vrugt}\ \emph {et~al.}(2022)\citenamefont {te~Vrugt}, \citenamefont {Holl}, \citenamefont {Koch}, \citenamefont {Wittkowski},\ and\ \citenamefont {Thiele}}]{te2022derivation}%
  \BibitemOpen
  \bibfield  {author} {\bibinfo {author} {\bibfnamefont {M.}~\bibnamefont {te~Vrugt}}, \bibinfo {author} {\bibfnamefont {M.~P.}\ \bibnamefont {Holl}}, \bibinfo {author} {\bibfnamefont {A.}~\bibnamefont {Koch}}, \bibinfo {author} {\bibfnamefont {R.}~\bibnamefont {Wittkowski}},\ and\ \bibinfo {author} {\bibfnamefont {U.}~\bibnamefont {Thiele}},\ }\href {https://doi.org/10.1088/1361-651X/ac856a} {\bibfield  {journal} {\bibinfo  {journal} {Model. Simul. Mater. Sci. Eng.}\ }\textbf {\bibinfo {volume} {30}},\ \bibinfo {pages} {084001} (\bibinfo {year} {2022})}\BibitemShut {NoStop}%
\bibitem [{\citenamefont {Frohoff-H\"ulsmann}\ \emph {et~al.}(2021)\citenamefont {Frohoff-H\"ulsmann}, \citenamefont {Wrembel},\ and\ \citenamefont {Thiele}}]{PhysRevE.103.042602}%
  \BibitemOpen
  \bibfield  {author} {\bibinfo {author} {\bibfnamefont {T.}~\bibnamefont {Frohoff-H\"ulsmann}}, \bibinfo {author} {\bibfnamefont {J.}~\bibnamefont {Wrembel}},\ and\ \bibinfo {author} {\bibfnamefont {U.}~\bibnamefont {Thiele}},\ }\href {https://doi.org/10.1103/PhysRevE.103.042602} {\bibfield  {journal} {\bibinfo  {journal} {Phys. Rev. E}\ }\textbf {\bibinfo {volume} {103}},\ \bibinfo {pages} {042602} (\bibinfo {year} {2021})}\BibitemShut {NoStop}%
\bibitem [{\citenamefont {Saha}\ \emph {et~al.}(2020)\citenamefont {Saha}, \citenamefont {Agudo-Canalejo},\ and\ \citenamefont {Golestanian}}]{PhysRevX.10.041009}%
  \BibitemOpen
  \bibfield  {author} {\bibinfo {author} {\bibfnamefont {S.}~\bibnamefont {Saha}}, \bibinfo {author} {\bibfnamefont {J.}~\bibnamefont {Agudo-Canalejo}},\ and\ \bibinfo {author} {\bibfnamefont {R.}~\bibnamefont {Golestanian}},\ }\href {https://doi.org/10.1103/PhysRevX.10.041009} {\bibfield  {journal} {\bibinfo  {journal} {Phys. Rev. X}\ }\textbf {\bibinfo {volume} {10}},\ \bibinfo {pages} {041009} (\bibinfo {year} {2020})}\BibitemShut {NoStop}%
\bibitem [{\citenamefont {John}\ and\ \citenamefont {B\"ar}(2005)}]{PhysRevLett.95.198101}%
  \BibitemOpen
  \bibfield  {author} {\bibinfo {author} {\bibfnamefont {K.}~\bibnamefont {John}}\ and\ \bibinfo {author} {\bibfnamefont {M.}~\bibnamefont {B\"ar}},\ }\href {https://doi.org/10.1103/PhysRevLett.95.198101} {\bibfield  {journal} {\bibinfo  {journal} {Phys. Rev. Lett.}\ }\textbf {\bibinfo {volume} {95}},\ \bibinfo {pages} {198101} (\bibinfo {year} {2005})}\BibitemShut {NoStop}%
\bibitem [{\citenamefont {Frohoff-Hülsmann}\ and\ \citenamefont {Thiele}(2021)}]{10.1093/imamat/hxab026}%
  \BibitemOpen
  \bibfield  {author} {\bibinfo {author} {\bibfnamefont {T.}~\bibnamefont {Frohoff-Hülsmann}}\ and\ \bibinfo {author} {\bibfnamefont {U.}~\bibnamefont {Thiele}},\ }\href {https://doi.org/10.1093/imamat/hxab026} {\bibfield  {journal} {\bibinfo  {journal} {IMA J. Appl. Math.}\ }\textbf {\bibinfo {volume} {86}},\ \bibinfo {pages} {924} (\bibinfo {year} {2021})}\BibitemShut {NoStop}%
\bibitem [{\citenamefont {Frohoff-H\"ulsmann}\ \emph {et~al.}(2023)\citenamefont {Frohoff-H\"ulsmann}, \citenamefont {Holl}, \citenamefont {Knobloch}, \citenamefont {Gurevich},\ and\ \citenamefont {Thiele}}]{PhysRevE.107.064210}%
  \BibitemOpen
  \bibfield  {author} {\bibinfo {author} {\bibfnamefont {T.}~\bibnamefont {Frohoff-H\"ulsmann}}, \bibinfo {author} {\bibfnamefont {M.~P.}\ \bibnamefont {Holl}}, \bibinfo {author} {\bibfnamefont {E.}~\bibnamefont {Knobloch}}, \bibinfo {author} {\bibfnamefont {S.~V.}\ \bibnamefont {Gurevich}},\ and\ \bibinfo {author} {\bibfnamefont {U.}~\bibnamefont {Thiele}},\ }\href {https://doi.org/10.1103/PhysRevE.107.064210} {\bibfield  {journal} {\bibinfo  {journal} {Phys. Rev. E}\ }\textbf {\bibinfo {volume} {107}},\ \bibinfo {pages} {064210} (\bibinfo {year} {2023})}\BibitemShut {NoStop}%
\bibitem [{\citenamefont {Mandal}\ \emph {et~al.}(2024)\citenamefont {Mandal}, \citenamefont {Jaramillo},\ and\ \citenamefont {Sollich}}]{PhysRevE.109.L062602}%
  \BibitemOpen
  \bibfield  {author} {\bibinfo {author} {\bibfnamefont {R.}~\bibnamefont {Mandal}}, \bibinfo {author} {\bibfnamefont {S.~S.}\ \bibnamefont {Jaramillo}},\ and\ \bibinfo {author} {\bibfnamefont {P.}~\bibnamefont {Sollich}},\ }\href {https://doi.org/10.1103/PhysRevE.109.L062602} {\bibfield  {journal} {\bibinfo  {journal} {Phys. Rev. E}\ }\textbf {\bibinfo {volume} {109}},\ \bibinfo {pages} {L062602} (\bibinfo {year} {2024})}\BibitemShut {NoStop}%
\bibitem [{\citenamefont {Frohoff-H{\"u}lsmann}\ \emph {et~al.}(2023)\citenamefont {Frohoff-H{\"u}lsmann}, \citenamefont {Thiele},\ and\ \citenamefont {Pismen}}]{frohoff2023non}%
  \BibitemOpen
  \bibfield  {author} {\bibinfo {author} {\bibfnamefont {T.}~\bibnamefont {Frohoff-H{\"u}lsmann}}, \bibinfo {author} {\bibfnamefont {U.}~\bibnamefont {Thiele}},\ and\ \bibinfo {author} {\bibfnamefont {L.~M.}\ \bibnamefont {Pismen}},\ }\href {https://doi.org/10.1098/rsta.2022.0087} {\bibfield  {journal} {\bibinfo  {journal} {Phil. Trans. R. Soc. A}\ }\textbf {\bibinfo {volume} {381}},\ \bibinfo {pages} {20220087} (\bibinfo {year} {2023})}\BibitemShut {NoStop}%
\bibitem [{\citenamefont {Greve}\ \emph {et~al.}(2025)\citenamefont {Greve}, \citenamefont {Lovato}, \citenamefont {Frohoff-H{\"u}lsmann},\ and\ \citenamefont {Thiele}}]{GLFT2025prl}%
  \BibitemOpen
  \bibfield  {author} {\bibinfo {author} {\bibfnamefont {D.}~\bibnamefont {Greve}}, \bibinfo {author} {\bibfnamefont {G.}~\bibnamefont {Lovato}}, \bibinfo {author} {\bibfnamefont {T.}~\bibnamefont {Frohoff-H{\"u}lsmann}},\ and\ \bibinfo {author} {\bibfnamefont {U.}~\bibnamefont {Thiele}},\ }\href {https://doi.org/10.1103/PhysRevLett.134.018303} {\bibfield  {journal} {\bibinfo  {journal} {Phys. Rev. Lett.}\ }\textbf {\bibinfo {volume} {134}},\ \bibinfo {pages} {018303} (\bibinfo {year} {2025})}\BibitemShut {NoStop}%
\bibitem [{\citenamefont {Hohenberg}\ and\ \citenamefont {Halperin}(1977)}]{HoHa1977rmp}%
  \BibitemOpen
  \bibfield  {author} {\bibinfo {author} {\bibfnamefont {P.~C.}\ \bibnamefont {Hohenberg}}\ and\ \bibinfo {author} {\bibfnamefont {B.~I.}\ \bibnamefont {Halperin}},\ }\href {https://doi.org/10.1103/RevModPhys.49.435} {\bibfield  {journal} {\bibinfo  {journal} {Rev. Mod. Phys.}\ }\textbf {\bibinfo {volume} {49}},\ \bibinfo {pages} {435} (\bibinfo {year} {1977})}\BibitemShut {NoStop}%
\bibitem [{\citenamefont {Liu}\ \emph {et~al.}(2023)\citenamefont {Liu}, \citenamefont {Hou}, \citenamefont {Kitahata}, \citenamefont {He},\ and\ \citenamefont {Komura}}]{LHK2023jpsj}%
  \BibitemOpen
  \bibfield  {author} {\bibinfo {author} {\bibfnamefont {M.}~\bibnamefont {Liu}}, \bibinfo {author} {\bibfnamefont {Z.}~\bibnamefont {Hou}}, \bibinfo {author} {\bibfnamefont {H.}~\bibnamefont {Kitahata}}, \bibinfo {author} {\bibfnamefont {L.}~\bibnamefont {He}},\ and\ \bibinfo {author} {\bibfnamefont {S.}~\bibnamefont {Komura}},\ }\href {https://doi.org/10.7566/JPSJ.92.093001} {\bibfield  {journal} {\bibinfo  {journal} {J. Phys. Soc. Jpn.}\ }\textbf {\bibinfo {volume} {92}},\ \bibinfo {pages} {093001} (\bibinfo {year} {2023})}\BibitemShut {NoStop}%
\bibitem [{Note1()}]{Note1}%
  \BibitemOpen
  \bibinfo {note} {The local interaction is not equal to the total energy of the system, since the interaction energy for the two lattices $a$ and $b$ is different.}\BibitemShut {Stop}%
\bibitem [{\citenamefont {Glauber}(1963)}]{glauber_timedependent_1963}%
  \BibitemOpen
  \bibfield  {author} {\bibinfo {author} {\bibfnamefont {R.~J.}\ \bibnamefont {Glauber}},\ }\href {https://doi.org/10.1063/1.1703954} {\bibfield  {journal} {\bibinfo  {journal} {J. Math. Phys.}\ }\textbf {\bibinfo {volume} {4}},\ \bibinfo {pages} {294} (\bibinfo {year} {1963})}\BibitemShut {NoStop}%
\bibitem [{\citenamefont {Saito}\ and\ \citenamefont {Kubo}(1976)}]{Saito1976}%
  \BibitemOpen
  \bibfield  {author} {\bibinfo {author} {\bibfnamefont {Y.}~\bibnamefont {Saito}}\ and\ \bibinfo {author} {\bibfnamefont {R.}~\bibnamefont {Kubo}},\ }\href {https://doi.org/10.1007/BF01012879} {\bibfield  {journal} {\bibinfo  {journal} {J. Stat. Phys.}\ }\textbf {\bibinfo {volume} {15}},\ \bibinfo {pages} {233} (\bibinfo {year} {1976})}\BibitemShut {NoStop}%
\bibitem [{\citenamefont {Guislain}\ and\ \citenamefont {Bertin}(2024{\natexlab{d}})}]{PhysRevB.109.184203}%
  \BibitemOpen
  \bibfield  {author} {\bibinfo {author} {\bibfnamefont {L.}~\bibnamefont {Guislain}}\ and\ \bibinfo {author} {\bibfnamefont {E.}~\bibnamefont {Bertin}},\ }\href {https://doi.org/10.1103/PhysRevB.109.184203} {\bibfield  {journal} {\bibinfo  {journal} {Phys. Rev. B}\ }\textbf {\bibinfo {volume} {109}},\ \bibinfo {pages} {184203} (\bibinfo {year} {2024}{\natexlab{d}})}\BibitemShut {NoStop}%
\bibitem [{\citenamefont {Guislain}\ and\ \citenamefont {Bertin}(2023)}]{PhysRevLett.130.207102}%
  \BibitemOpen
  \bibfield  {author} {\bibinfo {author} {\bibfnamefont {L.}~\bibnamefont {Guislain}}\ and\ \bibinfo {author} {\bibfnamefont {E.}~\bibnamefont {Bertin}},\ }\href {https://doi.org/10.1103/PhysRevLett.130.207102} {\bibfield  {journal} {\bibinfo  {journal} {Phys. Rev. Lett.}\ }\textbf {\bibinfo {volume} {130}},\ \bibinfo {pages} {207102} (\bibinfo {year} {2023})}\BibitemShut {NoStop}%
\bibitem [{\citenamefont {Blom}(2023)}]{blom2023pair}%
  \BibitemOpen
  \bibfield  {author} {\bibinfo {author} {\bibfnamefont {K.}~\bibnamefont {Blom}},\ }\href {https://link.springer.com/book/10.1007/978-3-031-29612-3} {\emph {\bibinfo {title} {Pair-Correlation Effects in Many-Body Systems: Towards a Complete Theoretical Description of Pair-Correlations in the Static and Kinetic Description of Many-Body Systems}}}\ (\bibinfo  {publisher} {Springer Nature, Cham, Switzerland},\ \bibinfo {year} {2023})\BibitemShut {NoStop}%
\bibitem [{\citenamefont {Johnston}\ and\ \citenamefont {Plechác}(1998)}]{D_A_Johnston_1998SM}%
  \BibitemOpen
  \bibfield  {author} {\bibinfo {author} {\bibfnamefont {D.~A.}\ \bibnamefont {Johnston}}\ and\ \bibinfo {author} {\bibfnamefont {P.}~\bibnamefont {Plechác}},\ }\href {https://doi.org/10.1088/0305-4470/31/2/009} {\bibfield  {journal} {\bibinfo  {journal} {J. Phys. A: Math. Gen.}\ }\textbf {\bibinfo {volume} {31}},\ \bibinfo {pages} {475} (\bibinfo {year} {1998})}\BibitemShut {NoStop}%
\bibitem [{\citenamefont {Dhar}\ \emph {et~al.}(1997)\citenamefont {Dhar}, \citenamefont {Shukla},\ and\ \citenamefont {Sethna}}]{Deepak_Dhar_1997SM}%
  \BibitemOpen
  \bibfield  {author} {\bibinfo {author} {\bibfnamefont {D.}~\bibnamefont {Dhar}}, \bibinfo {author} {\bibfnamefont {P.}~\bibnamefont {Shukla}},\ and\ \bibinfo {author} {\bibfnamefont {J.~P.}\ \bibnamefont {Sethna}},\ }\href {https://doi.org/10.1088/0305-4470/30/15/013} {\bibfield  {journal} {\bibinfo  {journal} {J. Phys. A: Math. Gen.}\ }\textbf {\bibinfo {volume} {30}},\ \bibinfo {pages} {5259} (\bibinfo {year} {1997})}\BibitemShut {NoStop}%
\bibitem [{\citenamefont {Hanai}(2024)}]{PhysRevX.14.011029}%
  \BibitemOpen
  \bibfield  {author} {\bibinfo {author} {\bibfnamefont {R.}~\bibnamefont {Hanai}},\ }\href {https://doi.org/10.1103/PhysRevX.14.011029} {\bibfield  {journal} {\bibinfo  {journal} {Phys. Rev. X}\ }\textbf {\bibinfo {volume} {14}},\ \bibinfo {pages} {011029} (\bibinfo {year} {2024})}\BibitemShut {NoStop}%
\bibitem [{Note2()}]{Note2}%
  \BibitemOpen
  \bibinfo {note} {See Supplementary Material at [...]}\BibitemShut {NoStop}%
\bibitem [{\citenamefont {Strogatz}(2018)}]{strogatz2018nonlinear}%
  \BibitemOpen
  \bibfield  {author} {\bibinfo {author} {\bibfnamefont {S.~H.}\ \bibnamefont {Strogatz}},\ }\href@noop {} {\emph {\bibinfo {title} {Nonlinear dynamics and chaos: with applications to physics, biology, chemistry, and engineering}}}\ (\bibinfo  {publisher} {CRC press, Boca Raton, USA},\ \bibinfo {year} {2018})\BibitemShut {NoStop}%
\bibitem [{\citenamefont {Cross}\ and\ \citenamefont {Hohenberg}(1993)}]{RevModPhys.65.851}%
  \BibitemOpen
  \bibfield  {author} {\bibinfo {author} {\bibfnamefont {M.~C.}\ \bibnamefont {Cross}}\ and\ \bibinfo {author} {\bibfnamefont {P.~C.}\ \bibnamefont {Hohenberg}},\ }\href {https://doi.org/10.1103/RevModPhys.65.851} {\bibfield  {journal} {\bibinfo  {journal} {Rev. Mod. Phys.}\ }\textbf {\bibinfo {volume} {65}},\ \bibinfo {pages} {851} (\bibinfo {year} {1993})}\BibitemShut {NoStop}%
\bibitem [{\citenamefont {Dhooge}\ \emph {et~al.}(2003)\citenamefont {Dhooge}, \citenamefont {Govaerts},\ and\ \citenamefont {Kuznetsov}}]{10.1145/779359.779362}%
  \BibitemOpen
  \bibfield  {author} {\bibinfo {author} {\bibfnamefont {A.}~\bibnamefont {Dhooge}}, \bibinfo {author} {\bibfnamefont {W.}~\bibnamefont {Govaerts}},\ and\ \bibinfo {author} {\bibfnamefont {Y.~A.}\ \bibnamefont {Kuznetsov}},\ }\href {https://doi.org/10.1145/779359.779362} {\bibfield  {journal} {\bibinfo  {journal} {ACM Trans. Math. Softw.}\ }\textbf {\bibinfo {volume} {29}},\ \bibinfo {pages} {141–164} (\bibinfo {year} {2003})}\BibitemShut {NoStop}%
\bibitem [{\citenamefont {van~der Kolk}\ \emph {et~al.}(2023)\citenamefont {van~der Kolk}, \citenamefont {Garc\'{\i}a-P\'erez}, \citenamefont {Kouvaris}, \citenamefont {Serrano},\ and\ \citenamefont {Bogu\~n\'a}}]{PhysRevX.13.021038}%
  \BibitemOpen
  \bibfield  {author} {\bibinfo {author} {\bibfnamefont {J.}~\bibnamefont {van~der Kolk}}, \bibinfo {author} {\bibfnamefont {G.}~\bibnamefont {Garc\'{\i}a-P\'erez}}, \bibinfo {author} {\bibfnamefont {N.~E.}\ \bibnamefont {Kouvaris}}, \bibinfo {author} {\bibfnamefont {M.~A.}\ \bibnamefont {Serrano}},\ and\ \bibinfo {author} {\bibfnamefont {M.}~\bibnamefont {Bogu\~n\'a}},\ }\href {https://doi.org/10.1103/PhysRevX.13.021038} {\bibfield  {journal} {\bibinfo  {journal} {Phys. Rev. X}\ }\textbf {\bibinfo {volume} {13}},\ \bibinfo {pages} {021038} (\bibinfo {year} {2023})}\BibitemShut {NoStop}%
\bibitem [{\citenamefont {Anderson~Jr}\ and\ \citenamefont {Morley}(1985)}]{anderson1985eigenvalues}%
  \BibitemOpen
  \bibfield  {author} {\bibinfo {author} {\bibfnamefont {W.~N.}\ \bibnamefont {Anderson~Jr}}\ and\ \bibinfo {author} {\bibfnamefont {T.~D.}\ \bibnamefont {Morley}},\ }\href {https://doi.org/10.1080/03081088508817681} {\bibfield  {journal} {\bibinfo  {journal} {Linear and multilinear algebra}\ }\textbf {\bibinfo {volume} {18}},\ \bibinfo {pages} {141} (\bibinfo {year} {1985})}\BibitemShut {NoStop}%
\bibitem [{\citenamefont {Cohen}\ and\ \citenamefont {Havlin}(2010)}]{cohen2010complex}%
  \BibitemOpen
  \bibfield  {author} {\bibinfo {author} {\bibfnamefont {R.}~\bibnamefont {Cohen}}\ and\ \bibinfo {author} {\bibfnamefont {S.}~\bibnamefont {Havlin}},\ }\href@noop {} {\emph {\bibinfo {title} {Complex networks: structure, robustness and function}}}\ (\bibinfo  {publisher} {Cambridge university press},\ \bibinfo {year} {2010})\BibitemShut {NoStop}%
\bibitem [{\citenamefont {Blom}\ \emph {et~al.}(2023)\citenamefont {Blom}, \citenamefont {Ziethen}, \citenamefont {Zwicker},\ and\ \citenamefont {Godec}}]{PhysRevResearch.5.013135}%
  \BibitemOpen
  \bibfield  {author} {\bibinfo {author} {\bibfnamefont {K.}~\bibnamefont {Blom}}, \bibinfo {author} {\bibfnamefont {N.}~\bibnamefont {Ziethen}}, \bibinfo {author} {\bibfnamefont {D.}~\bibnamefont {Zwicker}},\ and\ \bibinfo {author} {\bibfnamefont {A.}~\bibnamefont {Godec}},\ }\href {https://doi.org/10.1103/PhysRevResearch.5.013135} {\bibfield  {journal} {\bibinfo  {journal} {Phys. Rev. Res.}\ }\textbf {\bibinfo {volume} {5}},\ \bibinfo {pages} {013135} (\bibinfo {year} {2023})}\BibitemShut {NoStop}%
\bibitem [{\citenamefont {Suchanek}\ \emph {et~al.}(2023)\citenamefont {Suchanek}, \citenamefont {Kroy},\ and\ \citenamefont {Loos}}]{PhysRevLett.131.258302}%
  \BibitemOpen
  \bibfield  {author} {\bibinfo {author} {\bibfnamefont {T.}~\bibnamefont {Suchanek}}, \bibinfo {author} {\bibfnamefont {K.}~\bibnamefont {Kroy}},\ and\ \bibinfo {author} {\bibfnamefont {S.~A.~M.}\ \bibnamefont {Loos}},\ }\href {https://doi.org/10.1103/PhysRevLett.131.258302} {\bibfield  {journal} {\bibinfo  {journal} {Phys. Rev. Lett.}\ }\textbf {\bibinfo {volume} {131}},\ \bibinfo {pages} {258302} (\bibinfo {year} {2023})}\BibitemShut {NoStop}%
\bibitem [{\citenamefont {Nardini}\ \emph {et~al.}(2017)\citenamefont {Nardini}, \citenamefont {Fodor}, \citenamefont {Tjhung}, \citenamefont {van Wijland}, \citenamefont {Tailleur},\ and\ \citenamefont {Cates}}]{PhysRevX.7.021007}%
  \BibitemOpen
  \bibfield  {author} {\bibinfo {author} {\bibfnamefont {C.}~\bibnamefont {Nardini}}, \bibinfo {author} {\bibfnamefont {E.}~\bibnamefont {Fodor}}, \bibinfo {author} {\bibfnamefont {E.}~\bibnamefont {Tjhung}}, \bibinfo {author} {\bibfnamefont {F.}~\bibnamefont {van Wijland}}, \bibinfo {author} {\bibfnamefont {J.}~\bibnamefont {Tailleur}},\ and\ \bibinfo {author} {\bibfnamefont {M.~E.}\ \bibnamefont {Cates}},\ }\href {https://doi.org/10.1103/PhysRevX.7.021007} {\bibfield  {journal} {\bibinfo  {journal} {Phys. Rev. X}\ }\textbf {\bibinfo {volume} {7}},\ \bibinfo {pages} {021007} (\bibinfo {year} {2017})}\BibitemShut {NoStop}%
\bibitem [{\citenamefont {Blom}\ \emph {et~al.}(2024)\citenamefont {Blom}, \citenamefont {Song}, \citenamefont {Vouga}, \citenamefont {Godec},\ and\ \citenamefont {Makarov}}]{blom2024milestoning}%
  \BibitemOpen
  \bibfield  {author} {\bibinfo {author} {\bibfnamefont {K.}~\bibnamefont {Blom}}, \bibinfo {author} {\bibfnamefont {K.}~\bibnamefont {Song}}, \bibinfo {author} {\bibfnamefont {E.}~\bibnamefont {Vouga}}, \bibinfo {author} {\bibfnamefont {A.}~\bibnamefont {Godec}},\ and\ \bibinfo {author} {\bibfnamefont {D.~E.}\ \bibnamefont {Makarov}},\ }\href {https://doi.org/10.1073/pnas.2318333121} {\bibfield  {journal} {\bibinfo  {journal} {Proc. Natl. Acad. Sci. U.S.A.}\ }\textbf {\bibinfo {volume} {121}},\ \bibinfo {pages} {e2318333121} (\bibinfo {year} {2024})}\BibitemShut {NoStop}%
\bibitem [{\citenamefont {Zhang}\ and\ \citenamefont {Garcia-Millan}(2023)}]{PhysRevResearch.5.L022033}%
  \BibitemOpen
  \bibfield  {author} {\bibinfo {author} {\bibfnamefont {Z.}~\bibnamefont {Zhang}}\ and\ \bibinfo {author} {\bibfnamefont {R.}~\bibnamefont {Garcia-Millan}},\ }\href {https://doi.org/10.1103/PhysRevResearch.5.L022033} {\bibfield  {journal} {\bibinfo  {journal} {Phys. Rev. Res.}\ }\textbf {\bibinfo {volume} {5}},\ \bibinfo {pages} {L022033} (\bibinfo {year} {2023})}\BibitemShut {NoStop}%
\bibitem [{\citenamefont {Frohoff-H\"ulsmann}\ and\ \citenamefont {Thiele}(2023)}]{PhysRevLett.131.107201}%
  \BibitemOpen
  \bibfield  {author} {\bibinfo {author} {\bibfnamefont {T.}~\bibnamefont {Frohoff-H\"ulsmann}}\ and\ \bibinfo {author} {\bibfnamefont {U.}~\bibnamefont {Thiele}},\ }\href {https://doi.org/10.1103/PhysRevLett.131.107201} {\bibfield  {journal} {\bibinfo  {journal} {Phys. Rev. Lett.}\ }\textbf {\bibinfo {volume} {131}},\ \bibinfo {pages} {107201} (\bibinfo {year} {2023})}\BibitemShut {NoStop}%
\bibitem [{\citenamefont {Greve}\ and\ \citenamefont {Thiele}(2024)}]{GrTh2024c}%
  \BibitemOpen
  \bibfield  {author} {\bibinfo {author} {\bibfnamefont {D.}~\bibnamefont {Greve}}\ and\ \bibinfo {author} {\bibfnamefont {U.}~\bibnamefont {Thiele}},\ }\href {https://doi.org/10.1063/5.0222013} {\bibfield  {journal} {\bibinfo  {journal} {Chaos}\ }\textbf {\bibinfo {volume} {34}},\ \bibinfo {pages} {123134} (\bibinfo {year} {2024})}\BibitemShut {NoStop}%
\bibitem [{\citenamefont {Penrose}(1991)}]{Penrose1991}%
  \BibitemOpen
  \bibfield  {author} {\bibinfo {author} {\bibfnamefont {O.}~\bibnamefont {Penrose}},\ }\href {https://doi.org/10.1007/bf01029993} {\bibfield  {journal} {\bibinfo  {journal} {J. Stat. Phys.}\ }\textbf {\bibinfo {volume} {63}},\ \bibinfo {pages} {975–986} (\bibinfo {year} {1991})}\BibitemShut {NoStop}%
\end{thebibliography}%


\begin{thebibliography}{10}%
\makeatletter
\providecommand \@ifxundefined [1]{%
 \@ifx{#1\undefined}
}%
\providecommand \@ifnum [1]{%
 \ifnum #1\expandafter \@firstoftwo
 \else \expandafter \@secondoftwo
 \fi
}%
\providecommand \@ifx [1]{%
 \ifx #1\expandafter \@firstoftwo
 \else \expandafter \@secondoftwo
 \fi
}%
\providecommand \natexlab [1]{#1}%
\providecommand \enquote  [1]{``#1''}%
\providecommand \bibnamefont  [1]{#1}%
\providecommand \bibfnamefont [1]{#1}%
\providecommand \citenamefont [1]{#1}%
\providecommand \href@noop [0]{\@secondoftwo}%
\providecommand \href [0]{\begingroup \@sanitize@url \@href}%
\providecommand \@href[1]{\@@startlink{#1}\@@href}%
\providecommand \@@href[1]{\endgroup#1\@@endlink}%
\providecommand \@sanitize@url [0]{\catcode `\\12\catcode `\$12\catcode `\&12\catcode `\#12\catcode `\^12\catcode `\_12\catcode `\%12\relax}%
\providecommand \@@startlink[1]{}%
\providecommand \@@endlink[0]{}%
\providecommand \url  [0]{\begingroup\@sanitize@url \@url }%
\providecommand \@url [1]{\endgroup\@href {#1}{\urlprefix }}%
\providecommand \urlprefix  [0]{URL }%
\providecommand \Eprint [0]{\href }%
\providecommand \doibase [0]{http://dx.doi.org/}%
\providecommand \selectlanguage [0]{\@gobble}%
\providecommand \bibinfo  [0]{\@secondoftwo}%
\providecommand \bibfield  [0]{\@secondoftwo}%
\providecommand \translation [1]{[#1]}%
\providecommand \BibitemOpen [0]{}%
\providecommand \bibitemStop [0]{}%
\providecommand \bibitemNoStop [0]{.\EOS\space}%
\providecommand \EOS [0]{\spacefactor3000\relax}%
\providecommand \BibitemShut  [1]{\csname bibitem#1\endcsname}%
\let\auto@bib@innerbib\@empty
%</preamble>
\bibitem [{\citenamefont {Avni}\ \emph {et~al.}(2023)\citenamefont {Avni}, \citenamefont {Fruchart}, \citenamefont {Martin}, \citenamefont {Seara},\ and\ \citenamefont {Vitelli}}]{SMavni2023nonreciprocal}%
  \BibitemOpen
  \bibfield  {author} {\bibinfo {author} {\bibfnamefont {Y.}~\bibnamefont {Avni}}, \bibinfo {author} {\bibfnamefont {M.}~\bibnamefont {Fruchart}}, \bibinfo {author} {\bibfnamefont {D.}~\bibnamefont {Martin}}, \bibinfo {author} {\bibfnamefont {D.}~\bibnamefont {Seara}}, \ and\ \bibinfo {author} {\bibfnamefont {V.}~\bibnamefont {Vitelli}},\ }\href@noop {} {\enquote {\bibinfo {title} {The non-reciprocal ising model},}\ } (\bibinfo {year} {2023}),\ \Eprint {http://arxiv.org/abs/2311.05471} {arXiv:2311.05471 [cond-mat.stat-mech]} \BibitemShut {NoStop}%
\bibitem [{\citenamefont {Glauber}(1963)}]{glauber_timedependent_1963SM}%
  \BibitemOpen
  \bibfield  {author} {\bibinfo {author} {\bibfnamefont {R.~J.}\ \bibnamefont {Glauber}},\ }\href {\doibase 10.1063/1.1703954} {\bibfield  {journal} {\bibinfo  {journal} {J. Math. Phys.}\ }\textbf {\bibinfo {volume} {4}},\ \bibinfo {pages} {294} (\bibinfo {year} {1963})}\BibitemShut {NoStop}%
\bibitem [{\citenamefont {Cross}\ and\ \citenamefont {Hohenberg}(1993)}]{SRevModPhys.65.851}%
  \BibitemOpen
  \bibfield  {author} {\bibinfo {author} {\bibfnamefont {M.~C.}\ \bibnamefont {Cross}}\ and\ \bibinfo {author} {\bibfnamefont {P.~C.}\ \bibnamefont {Hohenberg}},\ }\href {\doibase 10.1103/RevModPhys.65.851} {\bibfield  {journal} {\bibinfo  {journal} {Rev. Mod. Phys.}\ }\textbf {\bibinfo {volume} {65}},\ \bibinfo {pages} {851} (\bibinfo {year} {1993})}\BibitemShut {NoStop}%
\bibitem [{\citenamefont {Dhooge}\ \emph {et~al.}(2003)\citenamefont {Dhooge}, \citenamefont {Govaerts},\ and\ \citenamefont {Kuznetsov}}]{SM10.1145/779359.779362}%
  \BibitemOpen
  \bibfield  {author} {\bibinfo {author} {\bibfnamefont {A.}~\bibnamefont {Dhooge}}, \bibinfo {author} {\bibfnamefont {W.}~\bibnamefont {Govaerts}}, \ and\ \bibinfo {author} {\bibfnamefont {Y.~A.}\ \bibnamefont {Kuznetsov}},\ }\href {\doibase 10.1145/779359.779362} {\bibfield  {journal} {\bibinfo  {journal} {ACM Trans. Math. Softw.}\ }\textbf {\bibinfo {volume} {29}},\ \bibinfo {pages} {141–164} (\bibinfo {year} {2003})}\BibitemShut {NoStop}%
\bibitem [{\citenamefont {Yang}(1952)}]{SMPhysRev.85.808}%
  \BibitemOpen
  \bibfield  {author} {\bibinfo {author} {\bibfnamefont {C.~N.}\ \bibnamefont {Yang}},\ }\href {\doibase 10.1103/PhysRev.85.808} {\bibfield  {journal} {\bibinfo  {journal} {Phys. Rev.}\ }\textbf {\bibinfo {volume} {85}},\ \bibinfo {pages} {808} (\bibinfo {year} {1952})}\BibitemShut {NoStop}%
\bibitem [{\citenamefont {Mitrophanov}(2003)}]{Mitrophanov_2003}%
  \BibitemOpen
  \bibfield  {author} {\bibinfo {author} {\bibfnamefont {A.~Y.}\ \bibnamefont {Mitrophanov}},\ }\href {\doibase 10.1239/jap/1067436094} {\bibfield  {journal} {\bibinfo  {journal} {Journal of Applied Probability}\ }\textbf {\bibinfo {volume} {40}},\ \bibinfo {pages} {970–979} (\bibinfo {year} {2003})}\BibitemShut {NoStop}%
\bibitem [{\citenamefont {Bertoin}\ \emph {et~al.}(2004)\citenamefont {Bertoin}, \citenamefont {Martinelli},\ and\ \citenamefont {Peres}}]{bertoin2004lectures}%
  \BibitemOpen
  \bibfield  {author} {\bibinfo {author} {\bibfnamefont {J.}~\bibnamefont {Bertoin}}, \bibinfo {author} {\bibfnamefont {F.}~\bibnamefont {Martinelli}}, \ and\ \bibinfo {author} {\bibfnamefont {Y.}~\bibnamefont {Peres}},\ }\href@noop {} {\emph {\bibinfo {title} {Lectures on Probability Theory and Statistics: Ecole d'Ete de Probabilites de Saint-Flour XXVII-1997}}}\ (\bibinfo  {publisher} {Springer, Berlin Heidelberg, Germany},\ \bibinfo {year} {2004})\BibitemShut {NoStop}%
\bibitem [{\citenamefont {Martinelli}\ and\ \citenamefont {Olivieri}(1994)}]{martinelli1994approach}%
  \BibitemOpen
  \bibfield  {author} {\bibinfo {author} {\bibfnamefont {F.}~\bibnamefont {Martinelli}}\ and\ \bibinfo {author} {\bibfnamefont {E.}~\bibnamefont {Olivieri}},\ }\href@noop {} {\bibfield  {journal} {\bibinfo  {journal} {Communications in Mathematical Physics}\ }\textbf {\bibinfo {volume} {161}},\ \bibinfo {pages} {447} (\bibinfo {year} {1994})}\BibitemShut {NoStop}%
\bibitem [{\citenamefont {Johnston}\ and\ \citenamefont {Plechác}(1998)}]{D_A_Johnston_1998SM2}%
  \BibitemOpen
  \bibfield  {author} {\bibinfo {author} {\bibfnamefont {D.~A.}\ \bibnamefont {Johnston}}\ and\ \bibinfo {author} {\bibfnamefont {P.}~\bibnamefont {Plechác}},\ }\href {\doibase 10.1088/0305-4470/31/2/009} {\bibfield  {journal} {\bibinfo  {journal} {J. Phys. A: Math. Gen.}\ }\textbf {\bibinfo {volume} {31}},\ \bibinfo {pages} {475} (\bibinfo {year} {1998})}\BibitemShut {NoStop}%
\bibitem [{\citenamefont {Dhar}\ \emph {et~al.}(1997)\citenamefont {Dhar}, \citenamefont {Shukla},\ and\ \citenamefont {Sethna}}]{Deepak_Dhar_1997SM2}%
  \BibitemOpen
  \bibfield  {author} {\bibinfo {author} {\bibfnamefont {D.}~\bibnamefont {Dhar}}, \bibinfo {author} {\bibfnamefont {P.}~\bibnamefont {Shukla}}, \ and\ \bibinfo {author} {\bibfnamefont {J.~P.}\ \bibnamefont {Sethna}},\ }\href {\doibase 10.1088/0305-4470/30/15/013} {\bibfield  {journal} {\bibinfo  {journal} {J. Phys. A: Math. Gen.}\ }\textbf {\bibinfo {volume} {30}},\ \bibinfo {pages} {5259} (\bibinfo {year} {1997})}\BibitemShut {NoStop}%
\end{thebibliography}
\let\addcontentsline\oldaddcontentsline
%- --------------------------------
%- --------------------------------
\counterwithout{equation}{section}
\addtocounter{equation}{-1}

\clearpage
\newpage
\onecolumngrid
\renewcommand{\thefigure}{S\arabic{figure}}
\renewcommand{\theequation}{S\arabic{equation}}
\renewcommand{\thetable}{S\arabic{table}}
\renewcommand{\thesection}{S\arabic{section}}
\renewcommand{\thesubsection}{S\arabic{subsection}}
\setcounter{equation}{0}
\setcounter{table}{0}
\setcounter{figure}{0}
\setcounter{page}{1}
\setcounter{section}{0}

\begin{center}\textbf{Supplementary Material for: Local Order Controls the Onset of Oscillations in the Nonreciprocal Ising Model}\\[0.2cm]
Kristian Blom$^{1}$, Uwe Thiele$^{2,3,4}$, and Alja\v{z} Godec$^{1}$\\
\emph{$^{1}$Mathematical bioPhysics Group, Max Planck Institute for Multidisciplinary Sciences, 37077 G\"ottingen, Germany}\\
\emph{$^{2}$Institute of Theoretical Physics, University of M\"{u}nster, M\"{u}nster 48149, Germany}\\
\emph{$^{3}$Center for Nonlinear Science (CeNoS),
University of M\"{u}nster, M\"{u}nster 48149, Germany}\\
\emph{$^{4}$Center for Multiscale Theory and Computation (CMTC),
University of M\"{u}nster, M\"{u}nster 48149, Germany}\\[0.6cm]\end{center}

\begin{quotation}
In this Supplementary Material, we provide derivations and
further details of the results shown in the main paper. Furthermore, in
Sec.~\ref{proof} we prove that
static global order in the square-lattice Ising model persists
\emph{at least}
under small perturbations in the nonreciprocal coupling
$0<|K|\leq \mathcal{O}(1/N)$ for any finite but
  arbitrarily large system size $N\in
  [m,\infty)$ for some sufficiently large $m$.
\end{quotation}

%\hypersetup{allcolors=black}\tableofcontents\hypersetup{allcolors=mylinkcolor}
%---------------------------------
%---------------------------------
\section{The ``monomer'' approximation}
%---------------------------------
%---------------------------------
Here we introduce the monomer approximation to obtain an approximate expression for the probability $\mathcal{P}^{\mu}_{l,n}(t)$. The final results are also shown in Appendix F of the main paper. Suppose we want to know the probability of finding an up- or down-spin on the $a$ lattice that has $l$ up nearest neighbors on the $a$ lattice and $n$ nearest up neighbors on the $b$ lattice. On the monomer level we assume perfect mixing between the up and down spins, resulting in
\begin{equation}
    \mathcal{P}^{a }_{l,n}=
    \underbrace{\left[\binom{N^{a}_{+}}{l}\binom{N^{a}_{-}}{z-l}/\binom{N^{a}_{+}+N^{a}_{-}}{z}\right]}_{\substack{\rm probability \ for \\   \substack{l \ {\rm \  neighboring \ up} \\ {\rm spins \ on \ the} \ a \ {\rm lattice}}}}  \times  \underbrace{\left[\binom{N^{b}_{+}}{n}\binom{N^{b}_{-}}{1-n}/\binom{N^{b}_{+}+N^{b}_{-}}{1}\right]}_{\substack{\rm probability \ for \\   \substack{n \ {\rm \  neighboring \ up} \\ {\rm spins \ on \ the} \  b \ {\rm lattice}}}},
    \label{SMF1}
\end{equation}
where $N^{\mu}_{+}$ is the total number of up spins on the $\mu$ lattice, and similarly $N^{\mu}_{-}$ the total number of down spins. The same reasoning applies to the probability $\mathcal{P}^{b }_{l,n}$. Note that $N^{\mu}_{\pm}$ is generally time dependent, but for simplicity we omit the explicit time dependence. To relate $N^{\mu}_{\pm}$ to the global order we use the following relations
\begin{align}
    m^{\mu}&=(N^{\mu}_{+}-N^{\mu}_{-})/N, \\ 
    N&=N^{\mu}_{+}+N^{\mu}_{-}, 
\end{align}
from which follows that 
\begin{align}
    N^{\mu}_{\pm}&=N(1\pm m^{\mu})/2. \label{SrelN}
\end{align}
Inserting this back into Eq.~\eqref{SMF1} and taking the thermodynamic
limit, i.e.\ the scaling limit  $N \rightarrow \infty$ while keeping $\mathbf{m}(t)\equiv(m^{a}(t),m^{b}(t))$ fixed, we can make use of the following result for the binomial coefficients
\begin{equation}
    \lim\limits^{m^{\mu}={\rm const.}}_{N\rightarrow \infty }\binom{N(1\pm m^{\mu})/2}{l} \simeq \frac{(N(1\pm m^{\mu})/2)^{l}}{l!}, \ {\rm for} \ l \in \mathbb{N}
    \label{Sasympeq}
\end{equation}
where $\simeq$ stands for asymptotic equality. Inserting Eq.~\eqref{Sasympeq}  into Eq.~\eqref{SMF1}, and restoring the explicit time-dependence, we finally obtain (note that the $N$ dependence cancels out)
\begin{align}
    \mathcal{P}^{a }_{l,n}(t)
    &\simeq\frac{2\mathcal{C}^{z}_{l}(1+m^{a}(t))^{l}(1+m^{b}(t))^{n}}{(1-m^{a}(t))^{l-z}(1-m^{b}(t))^{n-1}}, \label{SPa} \\
    \mathcal{P}^{b}_{l,n}(t)
    &\simeq\frac{2\mathcal{C}^{z}_{l}(1+m^{b}(t))^{l}(1+m^{a}(t))^{n}}{(1-m^{b}(t))^{l-z}(1-m^{a}(t))^{n-1}},
    \label{SPb}
\end{align} 
where
\begin{equation}
    \mathcal{C}^{z}_{l}\equiv\frac{1}{2^{z+2}}\binom{z}{l}.
    \label{SC}
\end{equation}
Hence, in the monomer approximation, we find that the probability
$\mathcal{P}^{\mu }_{l,n}(t)$ only explicitly depends on the global
order $\mathbf{m}(t)$, and \emph{not} on the local order
$\mathbf{q}(t)\equiv(q^{aa}(t),q^{bb}(t),q^{ab}(t))$. 

\renewcommand{\thesubsection}{S1.\arabic{subsection}}
\subsection{Difference with respect to mean-field approximation}
Finally, let us point out the crucial difference between the mean-field approximation on 
the one hand and the
monomer and BG approximation on the other hand, which lies in the
treatment of the averaging of the term $\langle \tanh{\left(\Delta
  E^{\mu}_{i}/2\right)} \rangle $. Whereas the mean-field
approximation (uncontrollably) moves the average to the argument, i.e. $\langle \tanh{\left(\Delta E^{\mu}_{i}/2\right)} \rangle \approx \tanh{\left(\langle \Delta E^{\mu}_{i}/2 \rangle \right)}$,
the monomer and BG approximations use the fact that the value of $\Delta E^{\mu}_{i}/2$ lies in an enumerable set given by $U^{\mu}_{l,n}\equiv[2l-z]J_{\mu}+[2n-1]K_{\mu}$ with $l\in\{0,..,z\}$ and $n\in\{0,1\}$. This allows for an explicit summation
\begin{equation}
    \langle \tanh{\left(\Delta E^{\mu}_{i}/2\right)} \rangle =\sum_{l=0}^{z}\sum_{n=0}^{1}\mathcal{P}^{\mu}_{l,n}(t)\tanh{(U^{\mu}_{l,n})},
\end{equation}
where only the probability $\mathcal{P}^{\mu}_{l,n}(t)$ is approximated according to Eq.~\eqref{SPa}-\eqref{SPb} in the monomer approximation. 
%---------------------------------
%---------------------------------
%---------------------------------
\section{Linear stability analysis}
%---------------------------------
%---------------------------------
%---------------------------------
Here, we investigate the stability of steady-states of Eqs.~(8)-(10) in the main manuscript, in combination with monomer and Bethe-Guggenheim (BG) approximation, using linear stability analysis. This allows us to
identify the region in parameter space where we have a so-called Hopf
bifurcation, which marks the transition from a (non-oscillatory) steady-state to coherent oscillations.
%---------------------------------
%---------------------------------
\renewcommand{\thesubsection}{S2.\arabic{subsection}}
\subsection{The ``monomer'' approximation}
%---------------------------------
%---------------------------------
Since in the monomer approximation the local order is slaved by the global order, it suffices to consider the linear stability analysis for the global order. Let us consider a small perturbation of the global order around the steady-state value $\textbf{m}_{\rm s}$, i.e., $\mathbf{m}(t)=\mathbf{m}_{\rm s}+\mathbf{\delta m}(t)$. The most interesting steady-state value to consider is given by the disordered state $\mathbf{m}_{\rm s}=0$, in which case the probability \eqref{SPa}-\eqref{SPb} can be expanded as
\begin{align}
    \mathcal{P}^{a }_{l,n}(t)&=2\mathcal{C}^{z}_{l}\left(1+(2l-z)\delta m^{a}(t)+(2n-1)\delta m^{b}(t)\right)+\mathcal{O}(\mathbf{\delta m}^{2}(t)), \\
    \mathcal{P}^{b}_{l,n}(t)&=2\mathcal{C}^{z}_{l}\left(1+(2l-z)\delta m^{b}(t)+(2n-1)\delta m^{a}(t)\right)+\mathcal{O}(\mathbf{\delta m}^{2}(t)).
\end{align}
Inserting this linearized expression back into Eq.~(8) of the main manuscript, we eventually obtain
\begin{equation}
     \tau\frac{{\rm d} \mathbf{\delta m}(t)}{{\rm d}t}=\underbrace{\begin{pmatrix}
\hat{M}_{aa} & \hat{M}_{ab} \\
\hat{M}_{ba} & \hat{M}_{bb}
\end{pmatrix}}_{ \mathbf{\hat{M}}(J,K)} \mathbf{\delta m}(t)+\mathcal{O}(\mathbf{\delta m}^{2}(t)),
\label{Sm0MF}
\end{equation}
where the entries of the matrix $\mathbf{\hat{M}}$ read
\begin{align}
    \hat{M}_{\mu\mu}(J_{\mu},K_{\mu})&=2\sum_{l=0}^{z}\sum_{n=0}^{1}(2l-z)\mathcal{C}^{z}_{l}\tanh{(U^{\mu}_{l,n})}-1, \label{SSmf1} \\
    \hat{M}_{ab}(J_{a},K_{a})&=2\sum_{l=0}^{z}\sum_{n=0}^{1}(2n-1)\mathcal{C}^{z}_{l}\tanh{(U^{a}_{l,n})}, \\
    \hat{M}_{ba}(J_{b},K_{b})&=2\sum_{l=0}^{z}\sum_{n=0}^{1}(2n-1)\mathcal{C}^{z}_{l}\tanh{(U^{b}_{l,n})}.
    \label{SSmf2}
\end{align}
Note that $\hat{M}_{\mu\mu}$ can be written in terms of the steady-state solution for the local order, i.e., $\hat{M}_{\mu\mu}=zq^{\mu\mu}_{s}-1$, which we will make use of in the remaining calculation. Furthermore, the off-diagonals are related to the steady-state of the local order between the lattices, i.e., $\hat{M}_{ab}+\hat{M}_{ba}=2 q^{ab}_{s}$. Hence, a perturbation of the global order couples to the steady-state value of the local order. The solution of Eq.~\eqref{Sm0MF} can be written in terms of an eigenmode expansion
\begin{equation}
    \mathbf{\delta m}(t)=\sum_{k=\pm }\mathcal{A}_{k}\e{\hat{\lambda}_{k}t/\tau}\boldsymbol{\hat{\nu}}_{k},
\end{equation}
where $\mathcal{A}_{\pm}$ are set by the initial conditions, $\hat{\lambda}_{\pm}$ are the eigenvalues of the linear stability matrix 
\begin{equation}
    \hat{\lambda}_{\pm}=\left({\rm tr}(\mathbf{\hat{M}})\pm\sqrt{{\rm tr}(\mathbf{\hat{M}})^{2}-4{\rm det}(\mathbf{\hat{M}})}\right)/2,
    \label{SMFeigval}
\end{equation}
with 
\begin{align}
    {\rm tr}(\mathbf{\hat{M}})&=z\left(q^{aa}_{s}+q^{bb}_{s}\right)-2, \\
     {\rm det}(\mathbf{\hat{M}})&=(zq^{aa}_{s}-1)(zq^{bb}_{s}-1)-\hat{M}_{ab}\hat{M}_{ba},
\end{align}
and the eigenvectors $\boldsymbol{\hat{\nu}}_{\pm}$ read
\begin{equation}
    \boldsymbol{\hat{\nu}}_{\pm}=([\hat{\lambda}_{\pm}+2(1-zq^{bb}_{s})]/\hat{M}_{ba},1)^{\rm T}.
\end{equation}
Since all matrix entries in \eqref{Sm0MF} are real, the characteristic
polynomial also has real coefficients. Therefore, if the eigenvalues
are complex, they come in complex-conjugate pairs. The perturbation
$\mathbf{\delta m}(t)$ grows in time when ${\rm
  Re}(\hat{\lambda}_{\pm})>0$, and shrinks in time when ${\rm
  Re}(\hat{\lambda}_{\pm})<0$. The imaginary part of the eigenvalues
tells us whether the perturbation develops oscillations in time ${\rm
  Im}(\hat{\lambda}_{\pm})\neq 0$, or is monotonic in time ${\rm Im}(\hat{\lambda}_{\pm})=0$. The Hopf bifurcation, also known as an oscillatory instability or type-${\rm II}_{\rm o}$ instability in the Cross-Hohenberg classification \cite{SRevModPhys.65.851}, occurs when the complex conjugate eigenvalues cross the imaginary axis in the complex plane. Based on Eq.~\eqref{SMFeigval}, this occurs when  
\begin{align}
    {\rm tr}(\mathbf{\hat{M}})&=0 \rightarrow q^{aa}_{s}+q^{bb}_{s}=2/z
    \label{ScondMF1}, \\
    {\rm det}(\mathbf{\hat{M}})&>0 \rightarrow (zq^{aa}_{s}-1)(zq^{bb}_{s}-1)-\hat{M}_{ab}\hat{M}_{ba}>0.
    \label{ScondMF2}
\end{align}
For fixed $z$, both equations can be solved explicitly to obtain
expressions for the critical values of the parameters $J_{\mu}$ and $K_{\mu}$ on the line of Hopf bifurcations. Note that Eq.~\eqref{ScondMF1} sets a direct constraint on the
steady-state local order values. 
%---------------------------------
\renewcommand{\thesubsubsection}{S2.1.\arabic{subsubsection}}
\subsubsection{Perfectly nonreciprocal setting}
%---------------------------------
Focusing on the perfect nonreciprocal setting with $J_{a}=J_{b}=J$ and $K_{a}=-K_{b}=K$, we have $q^{aa}_{s}=q^{bb}_{s}\equiv q_{s}(J,K)$ and $\hat{M}_{ab}=-\hat{M}_{ba}$. Under these conditions, Eqs.~\eqref{ScondMF1}-\eqref{ScondMF2} transform into, 
\begin{align}
    {\rm tr}(\mathbf{\hat{M}})&=0 \rightarrow q_{s}=1/z, \label{ScondMF3} \\
    {\rm det}(\mathbf{\hat{M}})&>0 \rightarrow \hat{M}_{ab}\neq 0,
\end{align}
where the latter equation is directly satisfied for $K\neq0$. Hence, also in the ``monomer'' expression we find a critical value for the local order, which is given by $q_{\rm crit}\equiv 1/z$.
%---------------------------------
%---------------------------------
\subsection{The Bethe-Guggenheim ``pair'' approximation}
%---------------------------------
%---------------------------------
In the BG approximation, we can no longer neglect the local order for the linear stability analysis. Hence, we consider a small perturbation of the global and local order around their respective steady-state values
\begin{align}
    \mathbf{m}(t)&=\mathbf{m}_{s}+\delta \mathbf{m}(t), \\
    \mathbf{q}(t)&=\mathbf{q}_{s}+\delta \mathbf{q}(t).
\end{align}
For the sake of simplicity, we directly focus on the perfectly nonreciprocal setting with $J_{a}=J_{b}=J$ and $K_{a}=-K_{b}=K$. Upon inserting the steady-state values, the probability can be expanded up to first order in $\delta \mathbf{m}(t)$ and $\delta \mathbf{q}(t)$, which reduces Eqs.~(8)-(10) in the main paper to the following linear set of equations
\begin{equation}
    \tau \frac{{\rm d}}{{\rm d}t}\begin{pmatrix}
\delta \mathbf{ m}(t) \\
\delta \mathbf{q}(t)
\end{pmatrix}=\begin{pmatrix}
\mathbf{M} & \mathbf{0} \\
\mathbf{0} & \mathbf{Q}
\end{pmatrix}\begin{pmatrix}
\delta \mathbf{ m}(t) \\
\delta \mathbf{q}(t),
\end{pmatrix},
\end{equation}
where $\mathbf{M}(q_s;J,K)$ is a $2\times 2$ matrix and $\mathbf{Q}(q_s;J,K)$ a $3\times3$ matrix. Due to the diagonal block structure of the linearized equations, we can handle the perturbations for $\delta \mathbf{ m}(t)$ and $\delta \mathbf{q}(t)$ separately. The linear stability of $\delta \mathbf{ m}(t)$ is already discussed in the main paper, and here we proceed with $\delta \mathbf{q}(t)$. The elements of the $3\times3$ matrix $\mathbf{Q}$ are given by
\begin{align}
    Q_{11}&=Q_{22}=-2-\frac{2zq^{2}_{s}}{1-q^{2}_{s}}+\frac{2/z}{1-q^{2}_{s}}\sum_{l=0}^{z}\sum_{n=0}^{1}(2l-z)^{2}(\overline{\mathcal{P}}^{+}_{l}-\overline{\mathcal{P}}^{-}_{l})\tanh{(U^{a}_{l,n})}, \\
    Q_{12}&=Q_{21}=0,\\
    Q_{13}&=-Q_{23}=\frac{2}{z}\sum_{l=0}^{z}\sum_{n=0}^{1}(2l-z)(2n-1)(\overline{\mathcal{P}}^{+}_{l}-\overline{\mathcal{P}}^{-}_{l})\tanh{(U^{a}_{l,n})},\\
    Q_{31}&=-Q_{32}=\frac{1}{1-q^{2}_{s}}\sum_{l=0}^{z}\sum_{n=0}^{1}(2n-1)[(2l-z)(\overline{\mathcal{P}}^{+}_{l}-\overline{\mathcal{P}}^{-}_{l})-zq_{s}(\overline{\mathcal{P}}^{+}_{l}+\overline{\mathcal{P}}^{-}_{l})]\tanh{(U^{a}_{l,n})},\\
    Q_{33}&=-2+\sum_{\mu}\sum_{l=0}^{z}\sum_{n=0}^{1}(\overline{\mathcal{P}}^{+}_{l}-\overline{\mathcal{P}}^{-}_{l})\tanh{(U^{\mu}_{l,n})}. \label{Q33}
\end{align}
The solution of the linearized equation for $\delta \mathbf{q}(t)$ reads
\begin{equation}
    \delta \mathbf{q}(t)=\sum_{i=1}^{3}\mathcal{A}_{i}\e{\tilde{\lambda}_{i}t/\tau}\boldsymbol{\tilde{\nu}}_{i},
\end{equation}
where the $\mathcal{A}_{i}$ are determined by the initial conditions, and the eigenvalues of $\mathbf{Q}$, denoted as $\tilde{\lambda}_{i}$, are given by
\begin{align}
    \tilde{\lambda}_{1}&=Q_{11}, \\ 
    \tilde{\lambda}_{2}&=\frac{1}{2}\left(Q_{11}+Q_{33}+\sqrt{8Q_{13}Q_{31}+(Q_{11}-Q_{33})^{2}}\right), \\ 
    \tilde{\lambda}_{3}&=\frac{1}{2}\left(Q_{11}+Q_{33}-\sqrt{8Q_{13}Q_{31}+(Q_{11}-Q_{33})^{2}}\right),
\end{align}
and finally, the eigenvectors of $\mathbf{Q}$, denoted as $\boldsymbol{\tilde{\nu}}_{i}$, read
\begin{align}
    \boldsymbol{\tilde{\nu}}_{1}&=(1,1,0)^{\rm T}, \\ 
    \boldsymbol{\tilde{\nu}}_{2}&=(-Q_{13}/(\tilde{\lambda}_{2}-Q_{33}),Q_{13}/(\tilde{\lambda}_{2}-Q_{33}),1)^{\rm T}, \\ 
    \boldsymbol{\tilde{\nu}}_{3}&=(-Q_{13}/(\tilde{\lambda}_{3}-Q_{33}),Q_{13}/(\tilde{\lambda}_{3}-Q_{33}),1)^{\rm T}.
\end{align}
Note that $\tilde{\lambda}_{1}$ is always real, and therefore cannot give rise
to a Hopf bifurcation. The second and third eigenvalues can become complex, in which case their real part is given by ${\rm
  Re}(\tilde{\lambda}_{2,3})=(Q_{11}+Q_{33})/2$. However, we now prove that $Q_{11}\leq0$ and $Q_{33}\leq0$, and therefore also $\tilde{\lambda}_{2,3}$ cannot give rise to a Hopf bifurcation. 
  
  To prove that $Q_{11}\leq0$ we proceed with the following chain of inequalities
\begin{align}
    Q_{11}&=-2-\frac{2zq^{2}_{s}}{1-q^{2}_{s}}+\frac{2/z}{1-q^{2}_{s}}\sum_{l=0}^{z}\sum_{n=0}^{1}(2l-z)^{2}(\overline{\mathcal{P}}^{+}_{l}-\overline{\mathcal{P}}^{-}_{l})\tanh{(U^{a}_{l,n})} \nonumber \\
    &\leq -2-\frac{2zq^{2}_{s}}{1-q^{2}_{s}}+\Bigl |\frac{2/z}{1-q^{2}_{s}}\sum_{l=0}^{z}\sum_{n=0}^{1}(2l-z)^{2}(\overline{\mathcal{P}}^{+}_{l}-\overline{\mathcal{P}}^{-}_{l})\tanh{(U^{a}_{l,n})} \Bigr | \nonumber \\
    &\leq -2-\frac{2zq^{2}_{s}}{1-q^{2}_{s}}+\frac{2/z}{1-q^{2}_{s}}\sum_{l=0}^{z}\sum_{n=0}^{1}|(2l-z)^{2}(\overline{\mathcal{P}}^{+}_{l}-\overline{\mathcal{P}}^{-}_{l})\tanh{(U^{a}_{l,n})}| \nonumber \\
     &\leq -2-\frac{2zq^{2}_{s}}{1-q^{2}_{s}}+\frac{2/z}{1-q^{2}_{s}}\sum_{l=0}^{z}|(2l-z)^{2}(\overline{\mathcal{P}}^{+}_{l}-\overline{\mathcal{P}}^{-}_{l})||\tanh{([2l-z]J+K)}+\tanh{([2l-z]J-K)}| \nonumber \\
    &\leq -2-\frac{2zq^{2}_{s}}{1-q^{2}_{s}}+\frac{4/z}{1-q^{2}_{s}}\sum_{l=0}^{z}(2l-z)^{2}|\overline{\mathcal{P}}^{+}_{l}-\overline{\mathcal{P}}^{-}_{l}| \nonumber \\
   &\leq -2-\frac{2zq^{2}_{s}}{1-q^{2}_{s}}+\frac{4/z}{1-q^{2}_{s}}\sum_{l=0}^{z}(2l-z)^{2}(\overline{\mathcal{P}}^{+}_{l}+\overline{\mathcal{P}}^{-}_{l}) = 0,
\end{align}  
  where the last inequality follows from the triangle inequality $|\overline{\mathcal{P}}^{+}_{l}-\overline{\mathcal{P}}^{-}_{l}|\leq |\overline{\mathcal{P}}^{+}_{l}|+|\overline{\mathcal{P}}^{-}_{l}|$ together with  $\overline{\mathcal{P}}^{\pm}_{l}\geq 0$ and therefore $|\overline{\mathcal{P}}^{\pm}_{l}|=\overline{\mathcal{P}}^{\pm}_{l}$. Next, we proceed with $Q_{33}$ in a similar fashion
\begin{align}
    Q_{33}&=-2+\sum_{\mu}\sum_{l=0}^{z}\sum_{n=0}^{1}(\overline{\mathcal{P}}^{+}_{l}-\overline{\mathcal{P}}^{-}_{l})\tanh{(U^{\mu}_{l,n})}\nonumber \\
    &=-2+2\sum_{l=0}^{z}(\overline{\mathcal{P}}^{+}_{l}-\overline{\mathcal{P}}^{-}_{l})\left[\tanh{([2l-z]J+K)}+\tanh{([2l-z]J-K)}\right] \nonumber \\
    &\leq -2+  \Bigl |2\sum_{l=0}^{z}(\overline{\mathcal{P}}^{+}_{l}-\overline{\mathcal{P}}^{-}_{l})\left[\tanh{([2l-z]J+K)}+\tanh{([2l-z]J-K)}\right]  \Bigr | \nonumber \\
    &\leq -2+2\sum_{l=0}^{z}\Bigl |(\overline{\mathcal{P}}^{+}_{l}-\overline{\mathcal{P}}^{-}_{l})\left[\tanh{([2l-z]J+K)}+\tanh{([2l-z]J-K)}\right] \Bigr | \nonumber \\
    &\leq-2+4\sum_{l=0}^{z}|\overline{\mathcal{P}}^{+}_{l}-\overline{\mathcal{P}}^{-}_{l}| \nonumber \\
    &\leq-2+4\sum_{l=0}^{z}(\overline{\mathcal{P}}^{+}_{l}+\overline{\mathcal{P}}^{-}_{l})=-2+2=0,
\end{align}  
where for the last inequality we again used the triangle inequality.
This establishes that $Q_{11}\leq 0$ and $Q_{33}\leq 0$, and therefore when $\tilde{\lambda}_{2,3}$ become complex, their real part obeys the bound ${\rm
  Re}(\tilde{\lambda}_{2,3})=(Q_{11}+Q_{33})/2\leq 0$. Hence, up to first order, any perturbation
$\delta \mathbf{q}(t)$ decays over time. The \emph{coherent oscillations in
$\mathbf{q}(t)$} observed in Fig.~1c,d in the main paper \emph{are
therefore an inherently nonlinear effect} related to the coupling between
$\mathbf{m}(t)$ and $\mathbf{q}(t)$. 
%---------------------------------
%---------------------------------
%---------------------------------
\section{Proof of existence of static global order on the finite square lattice }\label{proof}
%---------------------------------
%---------------------------------
%---------------------------------
Here, we show that for any finite system size ($N<\infty$) there is at least a regime with $0<|K|\leq\mathcal{O}(1/N)$
  where the static global order in the two-dimensional square-lattice
  is \emph{not} destroyed. Moreover, in the thermodynamic scaling limit $N\rightarrow\infty$ and $|K|N={\rm constant}>0$ we show that there exists a regime for a nonzero magnetic field $h>0$ and for any coupling strength $J\geq J_{0}$ where the static global order must be preserved.
%---------------------------------
%---------------------------------
\renewcommand{\thesubsection}{S3.\arabic{subsection}}
\subsection{Recap of the global order}
%---------------------------------
%---------------------------------
For completeness, we recall some statements about the global order. We define the global order in lattice $a$ and $b$  as (here we explicitly write out the averaging $\langle \cdot \rangle$)
\begin{equation}
    m^{\mu}(t) \equiv N^{-1}\sum_{\{\boldsymbol{\sigma}\}}\sum_{i=1}^{N}\sigma^{\mu}_{i}P(\boldsymbol{\sigma};t),
    \label{m}
\end{equation}
where $\{\boldsymbol{\sigma}\}$ denotes the set of all possible spin configurations $\boldsymbol{\sigma}$. From the master equation (see Eq.~(2) in the main paper) we can obtain the time-evolution equation for the magnetization \cite{glauber_timedependent_1963SM}
\begin{equation}
    \frac{{\rm d} m^{\mu}(t)}{{\rm d}t}
    =-2N^{-1}\sum_{\{\boldsymbol{\sigma}\}}\sum_{i=1}^{N}\sigma^{\mu}_{i}w^{\mu}_{i}(\sigma^{\mu}_{i})P(\boldsymbol{\sigma};t).
\end{equation}
In the steady-state we have ${\rm d}m^{\mu}(t)/{\rm d}t=0$, and therefore
\begin{equation}
    \sum_{\{\boldsymbol{\sigma}\}}\sum_{i=1}^{N}\sigma^{\mu}_{i}w^{\mu}_{i}(\sigma^{\mu}_{i})P_{\rm s}(\boldsymbol{\sigma})=0,
    \label{ss}
\end{equation}
where $P_{\rm s}(\boldsymbol{\sigma})\equiv \lim_{t\rightarrow\infty}P(\boldsymbol{\sigma};t)$ denotes the steady-state probability. 
%---------------------------------
%---------------------------------
\subsection{Steady-state for $K=0$}
%---------------------------------
%---------------------------------
Let us first recall the steady-state global order for $K=0$, where we have two independent Ising systems on lattice $a$ and $b$, respectively. Let $P_{0, \rm s}(\boldsymbol{\sigma})\equiv \lim\limits_{t\rightarrow\infty}P(\boldsymbol{\sigma};t)|_{K=0}$ be the steady-state probability for $K=0$ and an arbitrary magnetic field $h$ (which we need later), and 
\begin{equation}
w_{0,i}^{\mu}(\sigma^{\mu}_{i})\equiv(1/2\tau)\left[1-\sigma^{\mu}_{i}\tanh{\left(J\sum\nolimits_{\langle i|j \rangle}\sigma^{\mu}_{j}+h\right)}\right]
\end{equation}
the transition rate when $K=0$ in the presence of a magnetic field $h$. The steady-state global order for $K=0$, denoted as $m^{\mu}_{0,\rm s}$, reads \cite{SMPhysRev.85.808}
 \begin{equation}
     m^{\mu}_{0,\rm s}\equiv \lim\limits_{h\rightarrow0^{\pm}}N^{-1}\sum_{\{\boldsymbol{\sigma}\}}\sum_{i=1}^{N}\sigma^{\mu}_{i}P_{0,\rm s}(\boldsymbol{\sigma}),
     \label{m0}
 \end{equation}
which is nonzero for $J\gtrsim\ln{(1+\sqrt{2})}/2$
   when $N\gg 1$ (for $N\rightarrow \infty$ the sign $\gtrsim$ changes to $>$). Note that the limit $h\rightarrow 0^{\pm}$ of the magnetic field  weakly breaks the $\mathbb{Z}_{2}$-symmetry, such that the steady-state magnetization does not correspond to the unstable value with $m^{\mu}_{0,s}=0$.
%---------------------------------
%---------------------------------
\subsection{Perturbation expansion in $K$}
%---------------------------------
%---------------------------------
We consider a perturbation of the steady-state under a small change in $K$ such that $|K|\ll1$. For small perturbations we can expand the transition rates as follows,
\begin{equation}
    w_{i}^{\mu}(\sigma^{\mu}_{i})=w_{0, i}^{\mu}(\sigma^{\mu}_{i})+K\delta w_{i}^{\mu}(\sigma^{\mu}_{i})+\mathcal{O}(K^{2}),
\end{equation}
where it follows from a Taylor expansion of Eq.~(3) in the main paper that 
\begin{equation}
    \delta w_{i}^{\mu}(\sigma^{\mu}_{i})=(1/2\tau)[1-2\delta_{\mu,b}]\sigma^{a}_{i}\sigma^{b}_{i}\sech^{2}{\left(J\sum\nolimits_{\langle i|j \rangle}\sigma^{\mu}_{j}+h\right)},\label{dw}
\end{equation}
with $\delta_{\mu,b}=1$ when $\mu=b$ and $\delta_{\mu,b}=0$ when $\mu=a$.
A small perturbation in the transition rates induces a perturbation in the steady-state probability
\begin{equation}
    P_{\rm s}(\boldsymbol{\sigma})=P_{0, \rm s}(\boldsymbol{\sigma})+K\delta P_{\rm s}(\boldsymbol{\sigma})+\mathcal{O}(K^{2}),
\end{equation}
which in turn results in a perturbation of the magnetization, $m^{\mu}_{\rm s} = m^{\mu}_{0, \rm s}+K\delta m^{\mu}_{\rm s}+\mathcal{O}(K^{2})$,  
where it follows from Eq.~\eqref{m0} that
\begin{equation}
    \delta m^{\mu}_{\rm s} \equiv \lim\limits_{h\rightarrow0^{\pm}}N^{-1}\sum_{\{\boldsymbol{\sigma}\}}\sum_{i=1}^{N}\sigma^{\mu}_{i}\delta P_{\rm s}(\boldsymbol{\sigma}).
    \label{dmdef}
\end{equation}
Our aim is to provide an upper and lower bound for $\delta m^{\mu}_{\rm s}$. 
%---------------------------------
%---------------------------------
\subsection{Bound on perturbations of global order}
%---------------------------------
%---------------------------------
We start with an upper bound for $\delta m^{\mu}_{\rm s}$, which goes as follows:
\begin{align}
    \delta m^{\mu}_{\rm s}&\leq \lim\limits_{h\rightarrow0^{\pm}}|N^{-1}\sum_{\{\boldsymbol{\sigma}\}}\sum_{i=1}^{N}\sigma^{\mu}_{i}\delta P_{\rm s}(\boldsymbol{\sigma})| \nonumber \\
    &\leq \lim\limits_{h\rightarrow0^{\pm}}N^{-1}\sum_{\{\boldsymbol{\sigma}\}}\sum_{i=1}^{N}|\sigma^{\mu}_{i}\delta P_{\rm s}(\boldsymbol{\sigma})| \nonumber \\
     &=  \lim\limits_{h\rightarrow0^{\pm}}N^{-1}\sum_{\{\boldsymbol{\sigma}\}}\sum_{i=1}^{N}|\sigma^{\mu}_{i}||\delta P_{\rm s}(\boldsymbol{\sigma})| \nonumber \\
     &= \lim\limits_{h\rightarrow0^{\pm}} \sum_{\{\boldsymbol{\sigma}\}}|\delta P_{\rm s}(\boldsymbol{\sigma})|.
    \label{res1}
\end{align}
In exactly the same way, we can also provide a lower bound
\begin{align}
    \delta m^{\mu}_{\rm s}&\geq \lim\limits_{h\rightarrow0^{\pm}} -|N^{-1}\sum_{\{\boldsymbol{\sigma}\}}\sum_{i=1}^{N}\sigma^{\mu}_{i}\delta P_{\rm s}(\boldsymbol{\sigma})| \nonumber \\
    &\geq \lim\limits_{h\rightarrow0^{\pm}} -N^{-1}\sum_{\{\boldsymbol{\sigma}\}}\sum_{i=1}^{N}|\sigma^{\mu}_{i}\delta P_{\rm s}(\boldsymbol{\sigma})| \nonumber \\
     &= \lim\limits_{h\rightarrow0^{\pm}} -N^{-1}\sum_{\{\boldsymbol{\sigma}\}}\sum_{i=1}^{N}|\sigma^{\mu}_{i}||\delta P_{\rm s}(\boldsymbol{\sigma})|  \nonumber \\
     &= \lim\limits_{h\rightarrow0^{\pm}} -\sum_{\{\boldsymbol{\sigma}\}}|\delta P_{\rm s}(\boldsymbol{\sigma})|
    \label{res2}
\end{align}
Hence, combining the upper and lower bound, we obtain
\begin{equation}
    |\delta m^{\mu}_{\rm s}| \leq \lim\limits_{h\rightarrow0^{\pm}} \sum_{\{\boldsymbol{\sigma}\}}|\delta P_{\rm s}(\boldsymbol{\sigma})|.
\end{equation}
This is our first main result. We are left with determining an upper bound for $\lim_{h\rightarrow0^{\pm}} \sum_{\{\boldsymbol{\sigma}\}}|\delta P_{\rm s}(\boldsymbol{\sigma})|$, for which we use the following theorem shown in \cite{Mitrophanov_2003} (Theorem 2.1):
\begin{theorem}
Let the Markov chain $X(t)$ with infinitesimal generator $\mathbf{A}$, i.e., ${\rm d}\mathbf{p}(t)/{\rm d}t=\mathbf{p}(t)\mathbf{A}$, be exponentially weakly ergodic; that is, for any normalized initial conditions $\mathbf{p}(t=0)$, and $\mathbf{p}^{\dagger}(t=0)$, and any $t\geq 0$, there exists a $b>0$ and $c>2$ such that
\begin{equation}
    \|\mathbf{p}(t) - \mathbf{p}^{\dagger}(t)\| \leq c {\rm e}^{-b t}, \ t\geq 0,
\end{equation}
where $\|\mathbf{p}\|=\sum_{i}|p_{i}|$ denotes the $l_{1}$-norm for vectors. Then, for perturbations to the infinitesimal generator, $\mathbf{A}+\mathbf{\hat{A}}$, the following bound takes place for the perturbed stationary probabilities $\mathbf{\hat{p}}_{\rm s}$:
\begin{equation}
    \|\mathbf{p}_{\rm s} - \mathbf{\hat{p}}_{\rm s}\| \leq \frac{1+\ln{(c/2)}}{b}\|\mathbf{\hat{A}}\|,
    \label{finite}
\end{equation}
where $\|\mathbf{\hat{A}}\|=\max_{i}\sum_{j}|\hat{A}_{ij}|$ is the subordinate norm for the perturbation matrix. 
\label{Theorem1}
\end{theorem}

Since it is known that the finite volume Glauber dynamics on $\mathbb{Z}^{d}$ is exponentially weakly ergodic (see Theorem 3.3 on page 117 in \cite{bertoin2004lectures} together with proposition 3.9 on page 124 or simply use Eq.~(3.15) in \cite{bertoin2004lectures}), we can directly use Theorem 1.  To translate the results from Theorem 1 into a bound for $|\delta m^{\mu}_{\rm s}|$, we first want to bound the subordinate norm for the perturbed infinitesimal generator. To do so, note that the perturbation to the transition rates obeys the following bound,
\begin{equation}
    |K\delta w_{i}^{\mu}(\sigma^{\mu}_{i})|=|(K/2\tau)[1-2\delta_{\mu,b}]\sigma^{a}_{i}\sigma^{b}_{i}\sech^{2}{(J\sum\nolimits_{\langle i|j \rangle}\sigma^{\mu}_{j}+h)}|\leq |K|/2\tau,\label{dw2}
\end{equation}
where we used that $|\sech^{2}(x)|\leq1$ for $x\in\mathbb{R}$. These perturbations enter the off-diagonal terms of $\mathbf{\hat{A}}$, so we have established that $\hat{A}_{ij}\leq |K|/2\tau$ for $i \neq j$. Since we consider single spin-flip dynamics, each row/column in $\mathbf{\hat{A}}$ has $2N$ nonzero entries excluding the diagonal entry, which is equal to $\hat{A}_{ii}=-\sum_{j}\hat{A}_{ij}$. Therefore, the diagonal term can also be bounded by $|\hat{A}_{ii}|\leq 2N \times |K|/2\tau$. Combining these results, we obtain the following bound for the subordinate norm of the perturbed infinitesimal generator,
\begin{equation}
    \|\mathbf{\hat{A}}\|\leq 2\times (2N)\times |K|/2\tau = 2N|K|/\tau.
\end{equation}
Finally, we note that $\|\mathbf{p}_{\rm s} - \mathbf{\hat{p}}_{\rm s}\|$ in Theorem 1 translates into $\lim_{h\rightarrow0^{\pm}}|K|\sum_{\{\boldsymbol{\sigma}\}}|\delta P_{\rm s}(\boldsymbol{\sigma})|$ in our work. Combining all together, we obtain that there exists $b>0$ and $c>2$ such that for $K\neq0$
\begin{equation}
\lim\limits_{h\rightarrow0^{\pm}}\sum_{\{\boldsymbol{\sigma}\}}|\delta P_{\rm s}(\boldsymbol{\sigma})|\leq \frac{2N(1+\ln{(c/2)})}{b\tau} ,
\label{bound}
\end{equation}
and therefore the perturbation is bounded by a finite number (when $N<\infty$ and $\tau\neq0$) independent of $K$ up to first order
\begin{equation}
    |\delta m^{\mu}_{\rm s}| \leq \frac{2N(1+\ln{(c/2))}}{b\tau} < \infty. 
    \label{mfinite}
\end{equation}
%---------------------------------
%---------------------------------
\subsection{Connecting the dots}
%---------------------------------
%---------------------------------
We have shown that up to first order the static global order can be written as $m^{\mu}_{\rm s} = m^{\mu}_{0, \rm s}+K\delta m^{\mu}_{\rm s}+\mathcal{O}(K^{2})$ under a small perturbation in $K$, where $m^{\mu}_{0, \rm s}\neq 0$ for $J\gtrsim\ln{(1+\sqrt{2})}/2$ as shown in Eq.~\eqref{m0}. Furthermore, the perturbation $\delta m^{\mu}_{\rm s}$ is bounded by a finite number independent of $K$, as shown in Eq.~\eqref{mfinite}. This means that for arbitrarily small nonzero $K$, the perturbed global order gets arbitrary close to the unperturbed value, i.e.
\begin{equation}
    |m^{\mu}_{\rm s}-m^{\mu}_{0, \rm s}|\leq |K| \left(\frac{2N(1+\ln{(c/2)})}{b\tau}\right).
    \label{bound2}
\end{equation}
To be more precise, if we set
\begin{equation}
    0<|K|\leq \mathcal{O}(1/N),
\end{equation}
the perturbed steady-state magnetization must come arbitrarily close to the unperturbed steady-state magnetization by merely changing the proportionality factor.
If the static global order vanished for any arbitrary nonzero $K$ on the square lattice, it would indicate that $m^{\mu}_{\rm s}$ is a discontinuous and non-differentiable function of $K$ for $J\gtrsim\ln{(1+\sqrt{2})}/2$, as it would suddenly jump from $m^{\mu}_{0, \rm s}\neq0$ to $m^{\mu}_{\rm s}=0$ in the limit $K\rightarrow0^{+}$.
This cannot be true, since we have established that $m^{\mu}_{\rm s}$
gets arbitrarily close to $m^{\mu}_{0, \rm s}$ up to first order in
$K$. Hence, the static global order is not destroyed for finite system
sizes ($N<\infty$) in the perfectly nonreciprocal Ising model at least
when $0<K\leq \mathcal{O}(1/N)$.

Note that our proof does not state anything about the regime with coherent steady-state oscillations on the square lattice. Similar to the results in \cite{SMavni2023nonreciprocal}, we also observe that spiral defects can destroy coherent steady-state oscillations on the square lattice. This, however, does not have any implication for the regime with static order, which is our main focus in this Section. 
%---------------------------------
%---------------------------------
\subsection{What happens in the thermodynamic limit?}
%---------------------------------
%---------------------------------
What happens to the static global order when we take $N\rightarrow\infty$? It is known that the infinite volume Glauber dynamics is not ergodic as the Gibbs measure is not unique \cite{bertoin2004lectures}. Therefore, one cannot use Theorem \ref{Theorem1}. However, in the presence of a magnetic field with $h>0$ there exists a region $J\geq J_{0}$ (this includes the infinite $J\rightarrow\infty$ limit) such that the infinite volume Glauber dynamics has exponential convergence, i.e., is weakly ergodic, as shown in Theorem 5.1b in \cite{martinelli1994approach} (page 479). Hence, in this regime  we can use Theorem \ref{Theorem1} to obtain a bound for $|K|\sum_{\{\boldsymbol{\sigma}\}}|\delta P_{\rm s}(\boldsymbol{\sigma})|$, which would also be given by Eq.~\eqref{bound} (albeit that $b$ and $c$ might change due to the presence of $h>0$). Looking at  Eq.~\eqref{bound2} we can then take a scaling limit $N\rightarrow \infty$ and $|K|N={\rm constant}$, such that the perturbed steady-state long-range order can get arbitrarily close to the unperturbed value by simply changing the proportionality constant. Hence, we conclude that there exists a regime for $|K|N={\rm constant}$ where the static order is preserved when
\begin{equation}
    N\xrightarrow[h>0,\ J\geq J_{0}]{|K|N ={\rm const.}}\infty.
\end{equation}
%---------------------------------
%---------------------------------
%---------------------------------
%---------------------------------
%---------------------------------
%---------------------------------
%

\end{document}